\documentclass[prc,aps,floatfix,groupedaddress,nofootinbib,showpacs,preprintnumbers,
amsmath,amssymb,amsfonts,widetable] {revtex4-1}
\usepackage{bm}
\usepackage{soul}
\usepackage{array}
\usepackage{lipsum}
\usepackage{relsize}
\usepackage{mathrsfs}
\usepackage{amssymb}
\usepackage{amsmath}
\usepackage{graphicx}
\usepackage[usenames]{color}
\usepackage{pslatex}

\def\Cevens{CE$\nu$NS\hspace{3pt}}

\newcommand{\FF}[1]{{F}_{\raisebox{-0.5pt}{\!\tiny #1}}}
\newcommand{\rhoX}[1]{{\rho}_{\raisebox{-3.50pt}{\!\tiny #1}}}

\begin{document}
\title{Electroweak probes of ground state densities} 
\author{Junjie Yang}
\author{Jesse A. Hernandez}
\author{J. Piekarewicz}
\affiliation{Department of Physics, Florida State University,
               Tallahassee, FL 32306, USA}
\date{\today}
\begin{abstract}

\noindent\textbf{Background:} Elastic electron scattering has been used for decades 
to paint the most accurate picture of the proton distribution in atomic nuclei. This stands 
in stark contrast to the neutron distribution that is traditionally probed using hadronic 
reactions that are hindered by large uncertainties in the reaction mechanism. Spurred
by new experimental developments, it is now possible to gain valuable insights into 
the neutron distribution using exclusively electroweak probes. 
\smallskip

\noindent\textbf{Purpose:} To assess the information content and complementarity 
of the following three electroweak experiments in constraining the neutron distribution 
of atomic nuclei: (a) parity violating elastic electron scattering, (b) coherent elastic 
neutrino nucleus scattering, and (c) elastic electron scattering of unstable nuclei.
\smallskip

\noindent\textbf{Methods:} Relativistic mean-field models informed by the 
properties of finite nuclei and neutron stars are used to compute ground state 
densities and form factors of a variety of nuclei. All the models follow the same 
fitting protocol, except for the assumed---and presently unknown---value of the 
neutron skin thickness of ${}^{208}$Pb. This enables one to tune the density
dependence of the symmetry energy without compromising the success in
reproducing well known physical observables. 
\smallskip

\noindent\textbf{Results:} We found that the ongoing PREX-II and upcoming CREX 
campaigns at Jefferson Lab will play a vital role in constraining the weak form factor 
of xenon and argon, liquid noble gases that are used for the detection of both neutrinos 
and dark matter particles.
\smallskip

\noindent\textbf{Conclusions:} Remarkable new advances in experimental physics have 
opened a new window into ground state densities of atomic nuclei using solely electroweak 
probes. The diversity and versatility of these experiments reveal powerful correlations that 
impose important nuclear structure constraints. In turn, these constraints provide quantitative 
theoretical uncertainties that are instrumental in searches for new physics and insights into
the behavior of dense matter.

\end{abstract}

\smallskip
\maketitle

Among the most basic properties of an atomic nucleus are its mass and its radius.
At a deeper level, one would like to understand how the underlying nuclear 
dynamics determines the spatial distribution of protons and neutrons in the nuclear
ground state. However, given the intrinsic quark substructure of the nucleon, neither 
the proton nor the neutron densities can be determined from experiment. Rather, it 
is the charge distributions---both electric and weak---that are the genuine physical 
observables that properly incorporate the finite nucleon size\,\cite{Horowitz:2012we}. 
In the particular case of the (electric) charge density, elastic electron scattering 
experiments pioneered by Hofstadter in the late 1950's\,\cite{Hofstadter:1956qs},
together with subsequent refinements\,\cite{DeJager:1987qc,Fricke:1995,Angeli:2013},
have painted the most accurate picture of the spatial charge distribution. For example, 
the charge radius of ${}^{208}$Pb is known to about 0.02\% or 
$R_{\rm ch}^{208}\!=\!5.5012(13)\,{\rm fm}$\,\cite{Angeli:2013}. Such
impressive experimental success will culminate with the commissioning of electron 
scattering facilities dedicated to map the charge distribution of short-lived 
isotopes\,\cite{Suda:2009zz}. Given that the charge distribution of an atomic nucleus 
is strongly dominated by the protons, elastic electron scattering provides a powerful
experimental tool for the determination of the ground-state proton density.

Unfortunately, mapping the experimental weak-charge distribution has not enjoyed 
the same success. The main difficulty stems from the need for electroweak probes 
that require the design of enormously challenging experiments, such as parity-violating 
electron scattering or elastic neutrino scattering. Yet these experiments provide the 
cleanest determination of neutron densities. Indeed, given that the weak-charge of the 
proton is strongly suppressed by the weak mixing angle, {\sl i.e.,} 
$Q_{\rm W}^{p}\!=\!1\!-\!4\sin^{2}\!\theta_{\rm W}\!=\!0.0719(45)$\,\cite{Androic:2018kni},
the weak-charge distribution of a nucleus is dominated by neutrons in a manner
similar to how protons dominate the electric-charge distribution. Of course, myriad of 
experiments have focused on the determination of neutron densities. Indeed, some
of the premier experimental facilities throughout the world were commissioned with  
the primary goal of mapping the neutron distribution of atomic nuclei throughout the
nuclear chart. Among the most widely used experimental techniques to map the 
neutron distribution  is elastic proton scattering at intermediate energies 
($\sim\!200\!-\!800\,{\rm MeV}$)\,\cite{Saudinos:1974ef,Ray:1992fj}. Given that 
protons couple strongly to both neutrons and protons in the nuclear target, the 
extraction of the neutron density often relies on prior knowledge of the proton 
density distribution, which is obtained from ``unfolding" the single nucleon form 
factors from the charge density measured using elastic electron scattering; 
for a recent implementation of this technique, see 
Refs.\,\cite{Zenihiro:2010zz,Zenihiro:2018rmz} and references contained therein.
Although such experimental efforts are valuable---especially with the commissioning 
of rare isotopes facilities around the world\,\cite{Balantekin:2014opa}---the determination 
of neutron densities using hadronic probes is plagued by significant model dependencies 
and uncontrolled approximations. Although high statistics is the hallmark of hadronic 
experiments, the cost for the high efficiency are large systematic uncertainties associated 
with the theoretical interpretation. For a recent review on the vast arsenal of experimental 
techniques devoted to map the neutron distribution of atomic nuclei and their
associated uncertainties see\,\cite{Thiel:2019tkm}.

In an effort to mitigate hadronic uncertainties, a concerted effort has been devoted
to the use of electroweak probes to determine neutron densities. These efforts have 
been inspired by the 30 year-old realization that parity-violating electron scattering 
(PVES) offers a uniquely clean probe of neutron densities that is free from strong-interaction 
uncertainties\,\cite{Donnelly:1989qs}. The pioneering Lead Radius EXperiment (PREX) 
at the Jefferson Laboratory (JLab) has fulfilled this vision by providing the first 
model-independent determination of the weak-charge form factor of ${}^{208}$Pb, 
albeit at a single value of the momentum transfer\,\cite{Abrahamyan:2012gp,Horowitz:2012tj}. 
The weak-charge form factor is connected to the associated spatial distribution 
by means of a Fourier transform. In particular, knowledge of the \emph{entire} weak-charge 
form factor would enable the determination of the neutron density in much the same way as 
the measurement of the charge form factor determines the proton density. 
At the time of this writing, the follow-up PREX-II campaign at JLab was already underway.
PREX-II  will improve on the original PREX by reaching a precision in the measured weak-charge 
form factor that will translate into a $\sim\!0.06$\,fm sensitivity on the neutron radius of 
${}^{208}$Pb. In turn, the brand new \emph{Calcium Radius EXperiment} (CREX) is 
scheduled to run immediately after PREX-II. CREX will measure the weak-charge form 
factor of ${}^{48}$Ca with a high-enough precision to allow a determination of its neutron
radius to $\sim\!0.02\!-\!0.03$\,fm\,\cite{CREX:2013}. Beyond JLab, the Mainz Energy 
recovery Superconducting Accelerator (MESA), envisioned to start operations by 2023,
will pave the way for an era of high-precision parity-violating 
experiments\,\cite{Becker:2018ggl}. Within the scope of the P2 experiment, aimed to 
measure the weak charge of the proton with an unprecedented precision of 1.5\%, the 
Mainz Radius EXperiment (MREX) will determine the neutron radius of ${}^{208}$Pb
to $0.03$\,fm, which represents a factor of two improvement relative to PREX-II. Finally, 
fruitful discussions have started on the physics case for a measurement of the weak 
charge of ${}^{12}$C at MESA\,\cite{UNAM:2019}.

Besides its intrinsic value as a fundamental nuclear-structure observable, knowledge of the 
neutron distribution provides a powerful bridge to a diversity of physical phenomena. For 
example, the neutron-skin thickness of heavy nuclei, defined as the difference between 
the neutron and proton root-mean-square radii $R_{\rm skin}\!\equiv\!R_{n}\!-\!R_{p}$, 
is strongly correlated to the slope of the symmetry energy at saturation 
density\,\cite{Brown:2000,Furnstahl:2001un,Centelles:2008vu,RocaMaza:2011pm}---a 
fundamental parameter of the equation of state that impacts the structure, composition, and 
cooling mechanism of neutron stars\,\cite{Horowitz:2000xj,Horowitz:2001ya,Carriere:2002bx,
Steiner:2004fi,Erler:2012qd,Chen:2014sca,Chen:2014mza}. In turn, important dynamical 
signatures observed in the collision of heavy ions are encoded in the density dependence of the 
symmetry energy\,\cite{Tsang:2004zz,Chen:2004si,Steiner:2005rd,Shetty:2007zg,Tsang:2008fd,
Li:2008gp,Tsang:2012se,Horowitz:2014bja}. Finally, neutron densities play an important role in 
atomic parity-violating experiments that could provide a portal to new physics\,\cite{Wood:1997zq}. As 
in any weak-interaction process, the signal is inherently small and hindered by uncertainties in 
both atomic and nuclear-structure theory. However, measuring ratios of parity-violating 
observables along isotopic chains mitigates the sensitivity to atomic theory\,\cite{Antypas:2018mxf}. 
As a result, nuclear-structure uncertainties, primarily in the form of neutron radii, remain the 
limiting factor in the search for new physics\,\cite{Fortson:1990zz,Pollock:1992mv,Chen:1993fw,
RamseyMusolf:1999qk,Sil:2005tg}.

Although there is little doubt that parity-violating electron scattering experiments provide one 
of the cleanest probes of neutron densities, the experimental challenges are enormous (see 
Sec.\,{\ref{sec:PVES}}). Two recent developments involving electroweak processes may 
help mitigate some of the challenges and have opened new avenues of inquiry into the same 
compelling physics\,\cite{Akimov:2017ade,Brown:2017}. Published within a week of each other, 
one paper reported the first observation of coherent elastic neutrino-nucleus scattering 
(\Cevens)\,\cite{Akimov:2017ade}, while the second paper proposed the difference in the 
charge radii of mirror nuclei as a complement to the neutron skin thickness\,\cite{Brown:2017}.
Shortly after the discovery of weak neutral currents in 1973, \Cevens was suggested as a 
mechanism with favorable cross sections that could impact a variety of astrophysical phenomena,
such as neutrino transport in core-collapse supernovae and neutron stars\,\cite{Freedman:1973yd}. 
\Cevens is ``favorable" because the resulting (coherent) cross section is proportional to the 
square of the weak charge of the nucleus, which is dominated by the neutron number. 
As such, \Cevens becomes a powerful tool in the determination of the weak form factor of 
the nucleus, at least at low momentum transfers where the process remains coherent. However, 
low momentum transfers produce hard to detect low-energy nuclear recoils, a fact that hindered 
experimental confirmation for over four decades\,\cite{Akimov:2017ade}. Whereas the first 
observation of \Cevens benefited enormously from the technology developed for dark-matter 
searches having a similar recoil signature, \Cevens could ultimately cripple direct dark-matter 
searches through an irreducible neutrino background, the so-called ``neutrino-floor". Thus, 
\Cevens has applications in nuclear structure, fundamental symmetries, dark-matter 
searches, and supernovae detection, among many others\,\cite{Akimov:2017ade}. 

The argument in favor of using the difference in the charge radii of mirror nuclei as a complement 
to the neutron skin thickness is both simple and elegant\,\cite{Brown:2017}: in the limit of exact 
charge symmetry, the neutron radius of a given nucleus ({\sl e.g.,} ${}^{48}$Ca) is identical to 
the proton radius of its mirror partner ({\sl i.e.,} ${}^{48}$Ni). In this particular example, charge
symmetry demands the strict equality between the neutron skin thickness of ${}^{48}$Ca and 
the difference in proton radii between ${}^{48}$Ni and ${}^{48}$Ca. If true, this would imply that 
the sophisticated machinery that has been developed over many decades to probe the nuclear 
charge distribution via (parity-conserving) elastic electron scattering could be brought to bear 
on this fundamental problem. Although the basic idea is appealing, charge symmetry is known 
to be broken due to both electromagnetic effects and quark-mass differences\,\cite{Miller:2006tv}. 
Thus, the utility of the above argument relies on whether the correlation survives in the face of 
charge-symmetry violations. Naturally, most of the work in Ref.\,\cite{Brown:2017} was devoted 
to show that the differences in the charge radii of mirror nuclei as predicted by a set of Skyrme 
functionals displays a strong correlation to the associated neutron skin thickness---and ultimately 
to the density dependence of the symmetry energy---even in the presence of Coulomb corrections. 
Shortly thereafter, these conclusions were validated in the context of both relativistic energy density
functionals\,\cite{Yang:2017vih} and microscopic approaches using chiral 
interactions\,\cite{Sammarruca:2017siq}. Moreover, given that various neutron-star properties are 
sensitive to the density dependence of the symmetry energy, a ``data-to-data relation" emerged 
between the difference in charge radii of mirror nuclei and the radius of low-mass neutron 
stars\,\cite{Yang:2017vih}.

The main goal of this paper is to connect three electroweak processes---parity-violating electron 
scattering, coherent elastic neutrino-nucleus scattering, and parity-conserving electron 
scattering---in the quest to determine ground-state neutron densities which would ultimately
lead to powerful constraints on the equation of state of neutron-rich matter. Although the connection 
is compelling, each challenging experiment brings its own strengths and weaknesses, and thus the 
need for a concerted effort.  With this goal in mind, the paper has been organized as follows. In 
Sec.\,\ref{sec:Formalism} we discuss briefly each of the relevant electroweak process and highlight
their role as clean and model-independent probes of ground-state densities. In turn, 
Sec.\,\ref{sec:Results} discusses predictions and correlations among various ground-state properties 
within the framework of covariant density functional theory. A summary of the main results and 
perspectives for future work are provided in Sec.\,\ref{sec:Conclusions}.


\section{Formalism}
\label{sec:Formalism}

In this section we describe briefly the three electroweak processes that are relevant to this work, 
their intimate connection to the neutron skin thickness, and ultimately to the equation of state of 
neutron rich matter.

\subsection{Elastic electron nucleus scattering}
\label{sec:ENS}

\begin{figure*}[ht]
\smallskip
 \includegraphics[width=0.4\columnwidth,height=4.75cm]{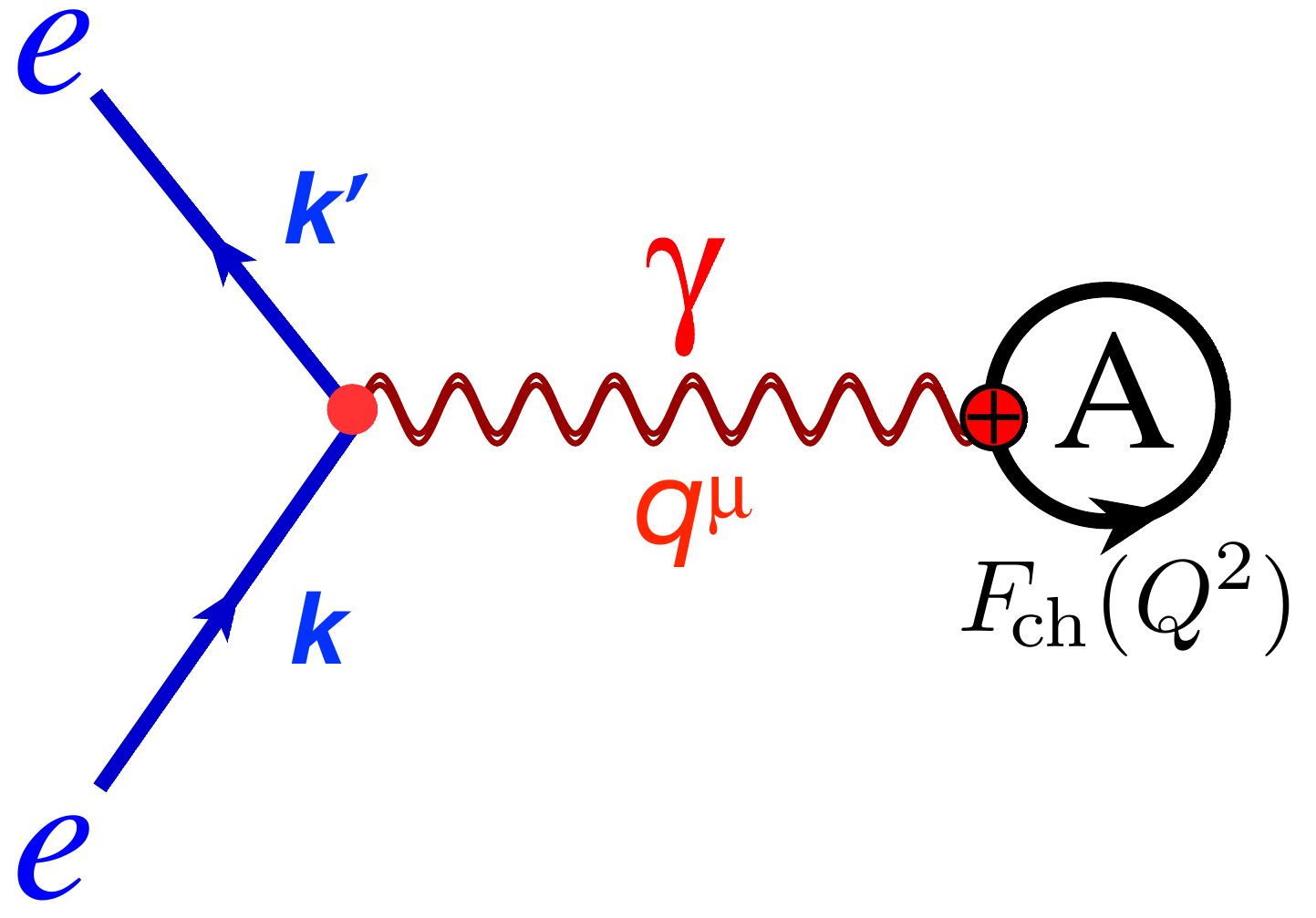}
\caption{(Color online) Feynman diagram for the elastic scattering of electrons from a spinless 
              nuclear target. Information on the internal structure of the nucleus is encoded in the 
              charge form factor $F_{\rm ch}(Q^{2})$.} 
\label{Fig1}
\end{figure*}

Mapping the charge distribution of atomic nuclei has been the source of intense experimental 
activity for over six decades\,\cite{Hofstadter:1956qs,DeJager:1987qc,Fricke:1995,Angeli:2013}.
Carried primarily---but not exclusively---by the protons,  the nuclear charge distribution has been
mapped with exquisite accuracy using elastic electron scattering. For the elastic scattering of an 
electron from a nuclear target, the Lorentz invariant matrix element may be written as follows:
\begin{equation}
  {\cal F}_{\!s'\!s}(Q^{2}) = \frac{e^{2}}{q^{2}}
  \Big[\overline{U}({\bf k}',s')\gamma_{\mu}U({\bf k},s)\Big]
  \langle p'|J_{\rm EM}^{\mu}|p\rangle,
 \label{EMME}
\end{equation}
where $k^{\mu}\!=\!(E,{\bf k})$ is the four-momentum of the incoming electron, $k^{\prime\mu}\!=\!(E',{\bf k}')$ 
the corresponding four-momentum of the scattered electron, and $q^{\mu}\!=\!k^{\mu}\!-\!k^{\prime\mu}$ the 
four-momentum transfer to the nucleus; see Fig.\ref{Fig1}.  For elastic scattering, the four-momentum carried 
by the photon may related to the nuclear momenta by energy-momentum conservation: 
$q^{\mu}\!=\!p^{\prime\mu}\!-\!p^{\mu}$. In the particular case of elastic scattering, the four-momentum transfer 
$q^{\mu}$ satisfies:
\begin{equation}
  q^{2}+2p\!\cdot\!q=0\; \stackrel{\rm lab}{\longrightarrow}\; Q^{2}=2M\omega,
 \label{ElKin}
\end{equation}
where $M$ is the mass of the target, $Q^{2}\!=\!-q^{2}\!>\!0$, and $\omega\!=\!q^{0}$ is the energy of the virtual 
photon as measured in the laboratory frame. 

In the particularly important case of elastic electron scattering from a spinless targets, the entire nuclear 
dynamics is encoded in a single Lorentz invariant \emph{charge form factor}. That is,
\begin{equation}
 \langle p'|J_{\rm EM}^{\mu}|p\rangle = ZF_{\rm ch}(Q^{2}) (p\!+\!p^{\prime})^{\mu},
 \label{ChFF}
\end{equation}
where $Z$ is the electric charge of the nucleus and the form factor has been normalized to one at zero 
momentum transfer. As a result, the Lorentz-invariant cross section may be written as follows:
\begin{equation}
   \frac{d\sigma}{dQ^{2}} = \frac{1}{4\pi}\left(\frac{e^{2}}{Q^{2}}\right)^{\!2}
   \left[\frac{(k\cdot p)(k'\cdot p)-M^{2}Q^{2}\!/4}{(k\cdot p)^{2}}\right]
   Z^{2}F_{\rm ch}^{2}(Q^{2}),
\label{dedtEM}
\end{equation}
or as commonly written in the laboratory frame\,\cite{Aitchinson:1982}
\begin{equation}
    \left(\frac{d\sigma}{d\Omega}\right)_{\rm\!\!EM} = 
    \left[\frac{\alpha^{2}\cos^{2}(\theta/2)}{4E^{2}\sin^{4}(\theta/2)}\!\!\left(\frac{E'}{E}\right)\right]
    Z^{2}F_{\rm ch}^{2}(Q^{2}),
\label{dedOEM}
\end{equation}
where $\alpha$ is the fine-structure constant. The expression in brackets is the Mott 
cross section which represents the scattering of a relativistic (massless) electron from 
a spinless and structureless target. This term is given exclusively in terms of kinematical
variables and the fine structure constant. Deviations from this structureless limit are
imprinted in the charge form factor of the nucleus. Given that the form factor may 
be viewed as the Fourier transform of the spatial distribution, elastic electron scattering 
has painted the most accurate picture of the distribution of charge in atomic 
nuclei\,\cite{Walecka:2001}.

\subsection{Coherent elastic neutrino nucleus scattering}
\label{sec:CEvNS}

\begin{figure*}[ht]
\smallskip
 \includegraphics[width=0.4\columnwidth,height=4.75cm]{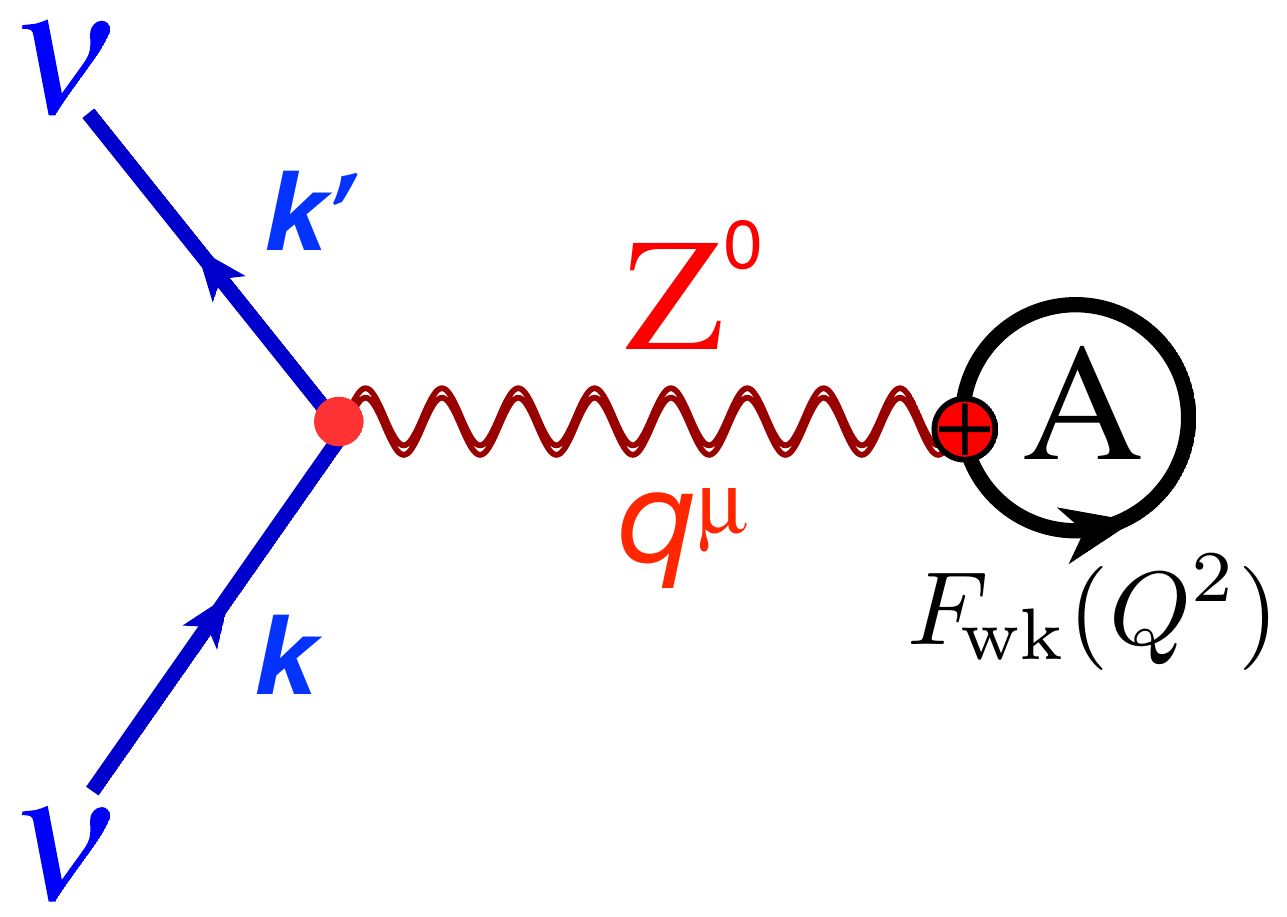}
\caption{(Color online) Feynman diagram for the elastic scattering of neutrinos from a spinless 
              nuclear target. Information on the internal structure of the nucleus is encoded in the 
              weak charge form factor $F_{\rm wk}(Q^{2})$.} 
\label{Fig2}
\end{figure*}

Unlike the long and successful history of elastic electron scattering as a sensitive probe 
of proton densities, no electroweak probe has been used effectively to map the neutron 
distribution---at least until very recently. Although the small weak charge of the proton 
makes elastic neutrino scattering a sensitive probe of neutron densities, the extremely
feeble weak interaction hinders the detection of the outgoing neutrino; see Fig.\,\ref{Fig2}. 
Thus, the detection of nuclear recoils of extremely low energy provides the sole 
alternative---an enormously challenging task that delayed the experimental confirmation 
of the coherent process\,\cite{Akimov:2017ade} by more than four decades since first 
suggested by Freedman\,\cite{Freedman:1973yd}.
Now that the coherent process has been observed, the application of \Cevens to the 
determination of neutron densities\,\cite{Patton:2012jr,Patton:2013nwa,Cadeddu:2017etk}
and supernovae detection has become a reality\,\cite{Horowitz:2003cz,Scholberg:2012id}.

In analogy to Eq.(\ref{EMME}), the Lorentz invariant matrix element for the elastic scattering 
of a neutrino from a nuclear target may be written as follows:
\begin{align}
  {\cal F}_{\!s'\!s}(Q^{2}) & =  \frac{M_{Z}^{2}}{16M_{W}^{2}}\frac{g^{2}}{Q^{2}+M_{Z}^{2}}
  \Big[\overline{U}({\bf k}',s')\gamma_{\mu}(1-\gamma^{5})U({\bf k},s)\Big]
  \langle p'|J_{\rm NC}^{\mu}|p\rangle \nonumber \\
  &  \hspace{25pt} \xrightarrow{Q^{2}\ll M_{Z}^{2}}\frac{G_{F}}{2\sqrt{2}}
  \Big[\overline{U}({\bf k}',s')\gamma_{\mu}(1-\gamma^{5})U({\bf k},s)\Big]
  \langle p'|J_{\rm NC}^{\mu}|p\rangle,
 \label{NCME}
\end{align}
where in the last line we have assumed that $Q^{2}\!\ll\!M_{Z}^{2}$, we have introduced the
dimensionful Fermi constant $G_{F}\!=\!g^{2}/4\sqrt{2}M_{W}^{2}$, and ``NC" stands for 
weak neutral current. If the neutrino scatters elastically from a spinless target, then as in the 
case of elastic electron scattering, the entire nuclear contribution to the reaction may be 
subsumed in a single Lorentz invariant \emph{weak form factor}. That is,
\begin{equation}
 \langle p'|J_{\rm NC}^{\mu}|p\rangle = Q_{\rm wk}F_{\rm wk}(Q^{2}) (p\!+\!p^{\prime})^{\mu},
 \label{ChFF}
\end{equation}
where $Q_{\rm wk}\!=\!-\!N\!+\!(1\!-\!4\sin^{2}\!\theta_{\rm W})Z$ is the weak nuclear charge 
and the form factor has been normalized to one at zero momentum transfer. As noted earlier,
because of the suppression of the weak charge of the proton, most of the weak charge of the
nucleus is carried by the neutrons.
After carrying out the customary contraction between the leptonic and hadronic tensors, one 
obtains the Lorentz-invariant cross section:
\begin{equation}
   \frac{d\sigma}{dQ^{2}} = \frac{G_{\!F}^{2}}{8\pi}
   \left[\frac{(k\cdot p)(k'\cdot p)-M^{2}Q^{2}\!/4}{(k\cdot p)^{2}}\right]
   Q_{\rm wk}^{2}F_{\rm wk}^{2}(Q^{2}),
\label{CEvens0}
\end{equation}
where $k$($k'$) is the four-momentum of the incoming(outgoing) neutrino and $p$ is the initial 
four momentum of the target nucleus of mass $M$. Finally, evaluating the above differential cross 
in the laboratory frame in terms of the kinetic energy $T$ of the recoiling nucleus, one 
obtains\,\cite{Scholberg:2005qs}
\begin{equation}
   \left(\frac{d\sigma}{dT}\right)_{\rm\!\!NC} = \frac{G_{\!F}^{2}}{8\pi} M
   \left[2 - 2 \frac{T}{E} - \frac{MT}{E^{2}} \right]Q_{\rm wk}^{2}F_{\rm wk}^{2}(Q^{2}),
\label{CEvens1}
\end{equation}
where $E$ is the incident neutrino energy and $Q^{2}\!=\!2MT$. Note that the differential cross section
at forward angles is proportional to the \emph{square} of the weak charge of the nucleus, namely,
$Q_{\rm wk}^{2}\!\approx\!N^{2}$. This is the hallmark of the coherent reaction and the main reason for the
identification of \Cevens as having favorable cross sections\,\cite{Freedman:1973yd}, even if it took more 
than four decades for its experimental realization\,\cite{Akimov:2017ade}.

\subsection{Parity violating electron scattering}
\label{sec:PVES}

In an innovative paper written three decades ago, Donnelly, Dubach, and Sick proposed the use of 
parity violating electron scattering (PVES) as a clean and model-independent probe of neutron 
densities\,\cite{Donnelly:1989qs}. Since then\,\cite{Beise:1990hc}, many of the experimental challenges 
have been met, leading to a mature and enormously successful PVES program at 
JLab\,\cite{Aniol:2005zf,Abrahamyan:2012gp,Androic:2018kni}. Moreover, the interest in measuring
the neutron distribution of heavy nuclei (specifically of ${}^{208}$Pb) was rekindled because of the 
impact that such a measurement could have in constraining the equation of state of neutron rich 
matter and ultimately the structure of neutron stars\,\cite{Horowitz:2000xj}.

\begin{figure*}[ht]
\smallskip
 \includegraphics[width=0.7\columnwidth]{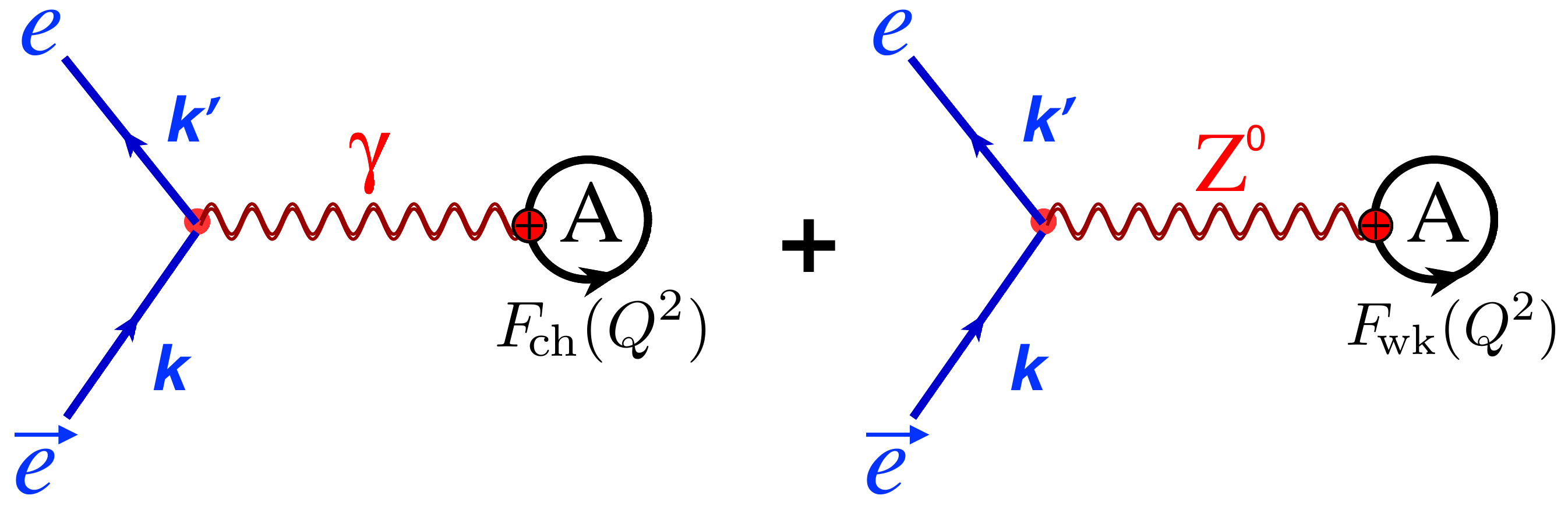}
\caption{(Color online) Feynman diagram for the elastic scattering of longitudinally polarized electrons 
              from a spinless nuclear target. In Born approximation, information on the internal structure of 
              the nucleus is encoded in the ratio between the weak $F_{\rm wk}(Q^{2})$ and the charge 
              $F_{\rm ch}(Q^{2})$ form  factors.} 
\label{Fig3}
\end{figure*}

As illustrated in Fig.\,\ref{Fig3}, a parity-violating asymmetry develops as a result of the quantum interference 
between two Feynman diagrams: a large one involving the exchange of a photon and a much smaller one 
involving the exchange of a $Z^{0}$ boson\,\cite{Musolf:1993tb,Walecka:2001}.
The parity-violating asymmetry is defined as 
 \begin{equation}
  A_{PV}= \frac{\displaystyle{\left(\frac{d\sigma}{d\Omega}\right)_{\!\!R}  - 
                                            \left(\frac{d\sigma}{d\Omega}\right)_{\!\!L}}}
                        {\displaystyle{\left(\frac{d\sigma}{d\Omega}\right)_{\!\!R}  + 
                        \left(\frac{d\sigma}{d\Omega}\right)_{\!\!L}}},
\label{APVa}
\end{equation}
where $(d\sigma\!/d\Omega)_{R/L}$ is the differential cross section for the elastic scattering of
right/left-handed longitudinally polarized electrons. Once electromagnetic and neutral-current 
matrix elements have been evaluated, as in Eqs.\,(\ref{EMME}) and (\ref{NCME}),  the cross 
section for right/left-handed electrons can be readily computed. One obtains,
\begin{equation}
 \left(\frac{d\sigma}{d\Omega}\right)_{\!\!R/L} \!\!= 
 \left(\frac{d\sigma}{d\Omega}\right)_{\rm\!\!EM} 
 \Big(1 \pm A_{PV}(Q^{2})\Big),
\label{dSigmaPol}
\end{equation}
where the parity-violating asymmetry is given by
\begin{equation}
  A_{PV}(Q^{2})  = \frac{G_{\!F}Q^{2}}{4\pi\alpha\sqrt{2}}
                              \frac{Q_{\rm wk}F_{\rm wk}(Q^{2})}{ZF_{ch}(Q^{2})}.
\label{APVb}
\end{equation}
We have continued to assume that the 4-momentum transfer is negligible as compared 
to the $Z^{0}$ mass ($Q^{2}\!\ll\!M_{\scriptscriptstyle{Z}}^{2}$) and Coulomb distortions 
have been ignored\,\cite{Horowitz:1998vv,RocaMaza:2008cg,RocaMaza:2011pm}. Note
that all information relevant to the \emph{spinless} nuclear target is encapsulated in the 
ratio between the weak and the charge form factors. Given that the distribution of charge 
is known accurately for many nuclei---certainly in the case of ${}^{48}$Ca and 
${}^{208}$Pb\,\cite{DeJager:1987qc}---measuring the parity violating asymmetry provides 
vital---and model-independent---information on the weak form factor. However, experiments
of this kind are enormously challenging because the scale of the asymmetry is set by the 
dimensionless ratio $Q^{2}/M_{W}^{2}$, which at the JLab kinematics is about $10^{-6}$.

\subsection{Symmetrized Fermi Function}
\label{sec:SFermi}

Although theoretical predictions will be provided in Sec.\ref{sec:Results} using a modern set of 
accurately calibrated relativistic density functionals\,\cite{Chen:2014mza}, these predictions are 
generated numerically. Often, a simple analytic form is desirable to compare against 
experimental results---especially at low momentum transfers where only a few moments of the 
density distribution are sufficient to constrain the experimental results. This is particularly true in 
the case of \Cevens or with experiments designed to probe physics beyond the standard 
model\,\cite{AristizabalSierra:2019zmy}. Assuming that the form factor $F\!(q)$ may be related to 
the spatial distribution $ \rho(r)$ through a Fourier transform, the low momentum-transfer 
behavior of the form factor may be written as follows: 
\begin{equation}
 F(q) = 1 - \frac{q^{2}}{3!}R^{2} + \frac{q^{4}}{5!}R^{4} - \frac{q^{6}}{7!}R^{6} + \ldots
 \label{Moments1}
\end{equation}
where $q\!\equiv\!\sqrt{Q^{2}}$ and the various moments of the spatial distribution are 
defined by
\begin{equation}
 R^{2n} \equiv \langle r^{2n} \rangle = \frac{\int r^{2n} \rho(r)d^{3}r}{\int \rho(r)d^{3}r}.
 \label{Moments2}
\end{equation}
Of particular interest is the mean square radius of the spatial distribution $R\!\equiv\!\sqrt{R^{2}}$, a quantity 
that in the case of the charge density often serves to calibrate energy density functionals\,\cite{Chen:2014sca}. 
As indicated above, the mean square radius may be computed from either the slope of the form factor at the 
origin or alternatively, from the second moment of the spatial distribution. 

In a recent publication we introduced\,\cite{Piekarewicz:2016vbn}, or rather \emph{re-introduced}\,\cite{Sprung:1997}, 
the symmetrized Fermi density defined by the following expression:
\begin{equation}
 {\large\rhoX{SF}}(r) \equiv {\large{\rhoX{0}}}\,\frac{\sinh(c/a)}{\cosh{(r/a)}+\cosh(c/a)}\; \hspace{4pt}{\rm where}\hspace{6pt}
 {\large\rhoX{0}}\equiv\frac{3A}{4\pi c\left(c^{2}+\pi^{2}a^{2}\right)} \; \hspace{4pt}{\rm and}\hspace{6pt} 
 \int\!{\large\rhoX{SF}}(r)d^{3}r=A.
 \label{RhoSF} 
\end{equation}
Although practically indistinguishable from the conventional and universally adopted Fermi density, the lesser 
known symmetrized version displays unique analytical properties that---unlike the conventional density---has 
a form factor that can be evaluated in closed analytic form\cite{Sprung:1997}: 
\begin{equation}
 \FF{SF}(q) = \frac{1}{A}\int e^{-i{\bf q}\cdot{\bf r}} {\large\rhoX{SF}}(r) d^{3}r 
             = \frac{3}{qc\Big((qc)^{2}+(\pi qa)^{2}\Big)}       
                  \left(\frac{\pi qa}{\sinh(\pi qa)}\right)
                  \left[\frac{\pi qa}{\tanh(\pi qa)}\sin(qc)-qc\cos(qc)\right]\,,
 \label{FFSF}                  
\end{equation}
with $\FF{SF}(q\!=\!0)\!=\!1.$ Among the many desirable features of an analytic form factor is that all the moments 
of the distribution can be evaluated exactly. That is, 
\begin{subequations}
\begin{align}
 R^{2}  & \equiv \langle r^{2} \rangle  = \frac{3}{5}c^{2} + 
  \frac{7}{5}(\pi a)^{2} \,,\\
 R^{4} & \equiv \langle r^{4} \rangle  = \frac{3}{7}c^{4} + 
  \frac{18}{7}(\pi a)^{2}c^{2} + \frac{31}{7}(\pi a)^{4}  \,,\\
 R^{6} & \equiv\langle r^{6} \rangle  = \frac{1}{3}c^{6} + 
  \frac{11}{3}(\pi a)^{2}c^{4} + \frac{239}{15}(\pi a)^{4}c^{2} + 
  \frac{127}{5}(\pi a)^{6} \,.
\end{align}
 \label{SFMoments}
\end{subequations}
Note that all these expressions are exact as they do not rely on a power series expansion in terms of the ``small'' 
parameter $\pi a/c$. Finally and highly insightful is the behavior of the symmetrized Fermi form factor in the limit 
of high momentum transfers:
\begin{equation}
 \FF{SF}(q) \rightarrow
 -6\frac{\pi a}{\sqrt{c^{2}+\pi^{2}a^{2}}}\frac{\cos(qc+\delta)}{qc}\,
 \mathlarger{e^{-\pi qa}}\,; \quad 
 \tan\delta\!\equiv\!\frac{\pi a}{c}\,.
  \label{HighqFs}
\end{equation}
This expression encapsulates many of the insights developed more than three decades ago in the context of the 
conventional Fermi function: diffractive oscillations controlled by the half-density radius $c$ and an exponential 
falloff driven by the diffuseness parameter $a$ (or rather $\pi a$)\,\cite{Amado:1979st,Amado:1986pm}. 

\section{Results}
\label{sec:Results}

We start this section by introducing the five relativistic mean field (RMF) models that will be used in this work. 
The models, falling under the general rubric of relativistic energy density functionals, are based on an underlying 
Lagrangian density containing an isodoublet nucleon field interacting via the exchange of various mesons and 
the photon\,\cite{Walecka:1974qa,Serot:1984ey}. Later on, various nonlinear terms were added to the Lagrangian 
density in an effort to improve the predictive power of the model\,\cite{Boguta:1977xi,Mueller:1996pm,
Horowitz:2000xj,Todd-Rutel:2005fa}. Once all physics insights (and biases) have been incorporated into the 
Lagrangian density, one proceeds to determine the parameters of the model by adopting a fitting protocol. In 
our case the calibration of the parameters is informed by ground-state properties of finite nuclei, their collective 
response, and constraints on the maximum neutron-star mass\,\cite{Chen:2014sca}. The outcome of the 
calibration procedure is an optimal set of parameters together with a covariance matrix that properly accounts 
for statistical uncertainties and correlations. The fitting protocol for all the models used in this paper is identical 
save one important distinction: the \emph{assumed} value for neutron skin thickness of 
${}^{208}$Pb ($R_{\rm skin}^{208}$). Indeed, the neutron skin thickness of ${}^{208}$Pb is allowed to vary 
over the relative wide range of $R_{\rm skin}^{208}\!=\!(0.12\!-\!0.32)\,{\rm fm}$\,\cite{Chen:2014sca,
Chen:2014mza}. The need to adopt such a prescription stems from the fact that the existent database of 
experimental observables is inadequate to constrain the isovector sector of the nuclear density functional---a 
fact that is reflected in our poor knowledge of the density dependence of the symmetry energy. By incorporating 
the neutron skin thickness of ${}^{208}$Pb---an observable recognized as a strong isovector 
indicator\,\cite{Brown:2000,Furnstahl:2001un,Centelles:2008vu,RocaMaza:2011pm}---one can successfully 
mitigate the problem.

\begin{figure*}[ht]
\smallskip
 \includegraphics[width=0.8\columnwidth]{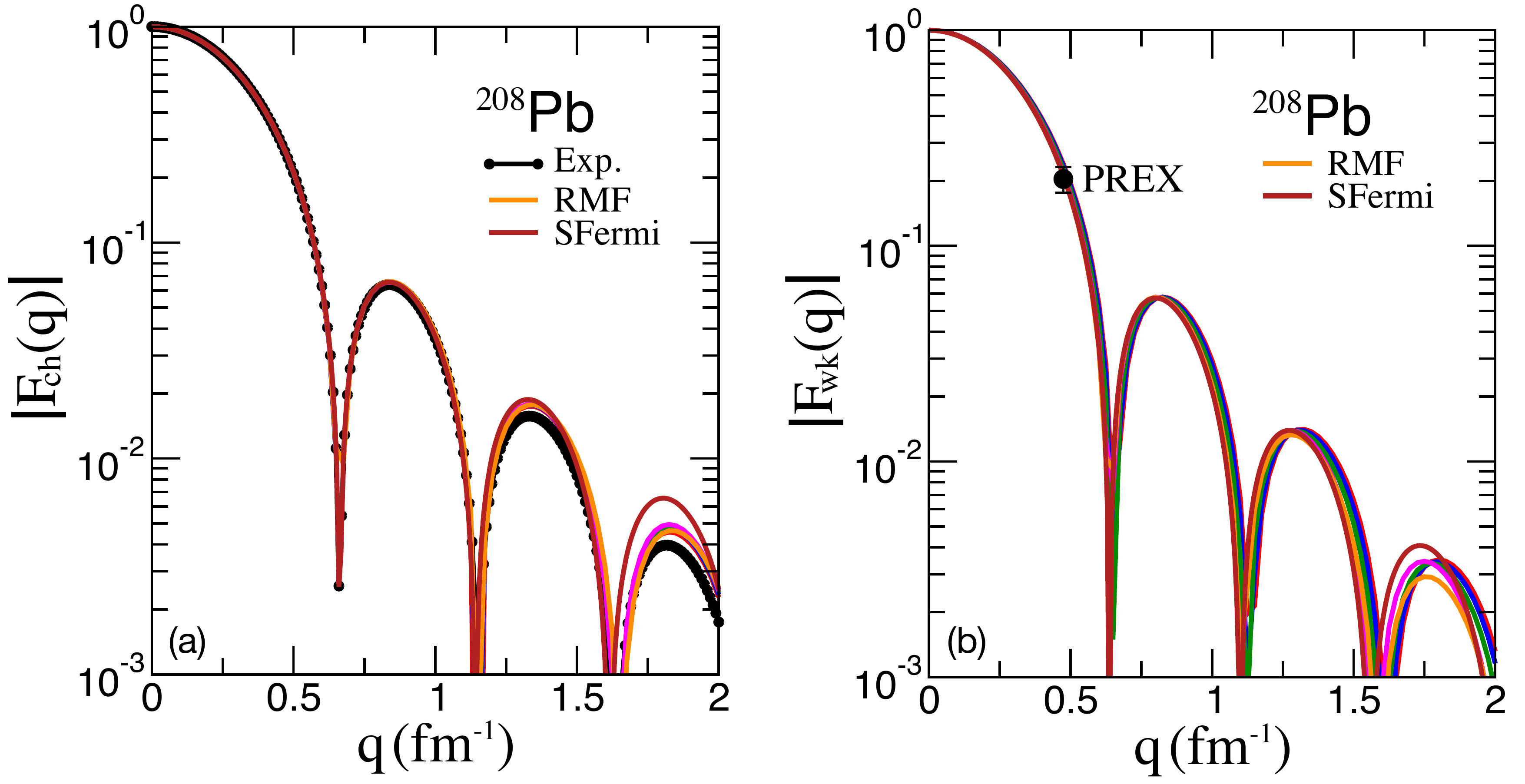}
\caption{(Color online) (a) The charge form factor of ${}^{208}$Pb as predicted by the set of 
relativistic mean-field models introduced in the text; the experimental data is from 
Ref.\,\cite{DeJager:1987qc}. Given that all models are informed by the charge radius of ${}^{208}$Pb, 
there are no visible differences in their predictions up to a momentum transfer of 
$q\!\lesssim\!2\,{\rm fm}^{-1}$. Predictions from a symmetrized Fermi function are also included to 
illustrate the characteristic diffractive oscillations and exponential falloff of the form factor; see 
Sec.\ref{sec:SFermi}. (b) The corresponding figure but now for the weak form factor of ${}^{208}$Pb. 
In this case the experimental information is reduced to the single point measured by the PREX
collaboration at the momentum transfer of $q\!=\!0.475\,{\rm fm}^{-1}$\,\cite{Abrahamyan:2012gp}.}
\label{Fig4}
\end{figure*}

In Fig.\ref{Fig4} we display predictions for the charge (left panel) and weak (right panel) form factors of 
${}^{208}$Pb as predicted by the five models calibrated in Ref.\,\cite{Chen:2014mza}. The experimental 
charge form factor is obtained from a Fourier-Bessel fit to the elastic electron scattering 
data\,\cite{DeJager:1987qc}. The theoretical band (labeled ``RMF") contains the predictions of all five 
models. Although all models are informed by the charge radius of ${}^{208}$Pb, the agreement with 
experiment extends well beyond the curvature at $q\!\equiv\!\sqrt{Q^{2}}\!=\!0$. An analytic symmetrized 
Fermi function fitted to the first two moments of the theoretical distribution is also included for comparison. 
The unmistakable diffractive oscillations and exponential falloff suggested by Eq.\,(\ref{HighqFs}) are clearly 
discernible in the figure. Moreover, although simple, the symmetrized Fermi function accurately reproduces 
the experimental cross section for nearly three diffraction minima. Unfortunately, the experimental situation 
concerning the weak form factor is diametrically opposed; see Fig.\ref{Fig4}(b). Indeed, the determination 
of the weak form factor of ${}^{208}$Pb at the single momentum transfer of $q\!=\!0.475\,{\rm fm}^{-1}$ by 
the PREX collaboration represents the sole electroweak measurement available today\,\cite{Abrahamyan:2012gp,
Horowitz:2012tj}. Although countless hadronic experiments that probe the neutron distribution have been 
conducted for decades, they are plagued---unlike electroweak measurements---by considerable model 
dependencies and uncontrolled approximations\,\cite{Thiel:2019tkm}. Alongside the sole experimental
point, predictions are displayed for the relativistic mean field models and a symmetrized Fermi function. 
Although difficult to appreciate using a logarithmic scale, the PREX error bar is simply too large to 
discriminate among the various theoretical models, even when their weak (or neutron) radii differ by 
about 3\%. Indeed, given that model dependences become discernible only at large momentum transfers,
and these high-momentum components dictate the structure of the density in the nuclear interior, it
is preferable to examine the spatial distribution of both electric and weak charge. 

\begin{figure*}[ht]
\smallskip
 \includegraphics[width=0.9\columnwidth]{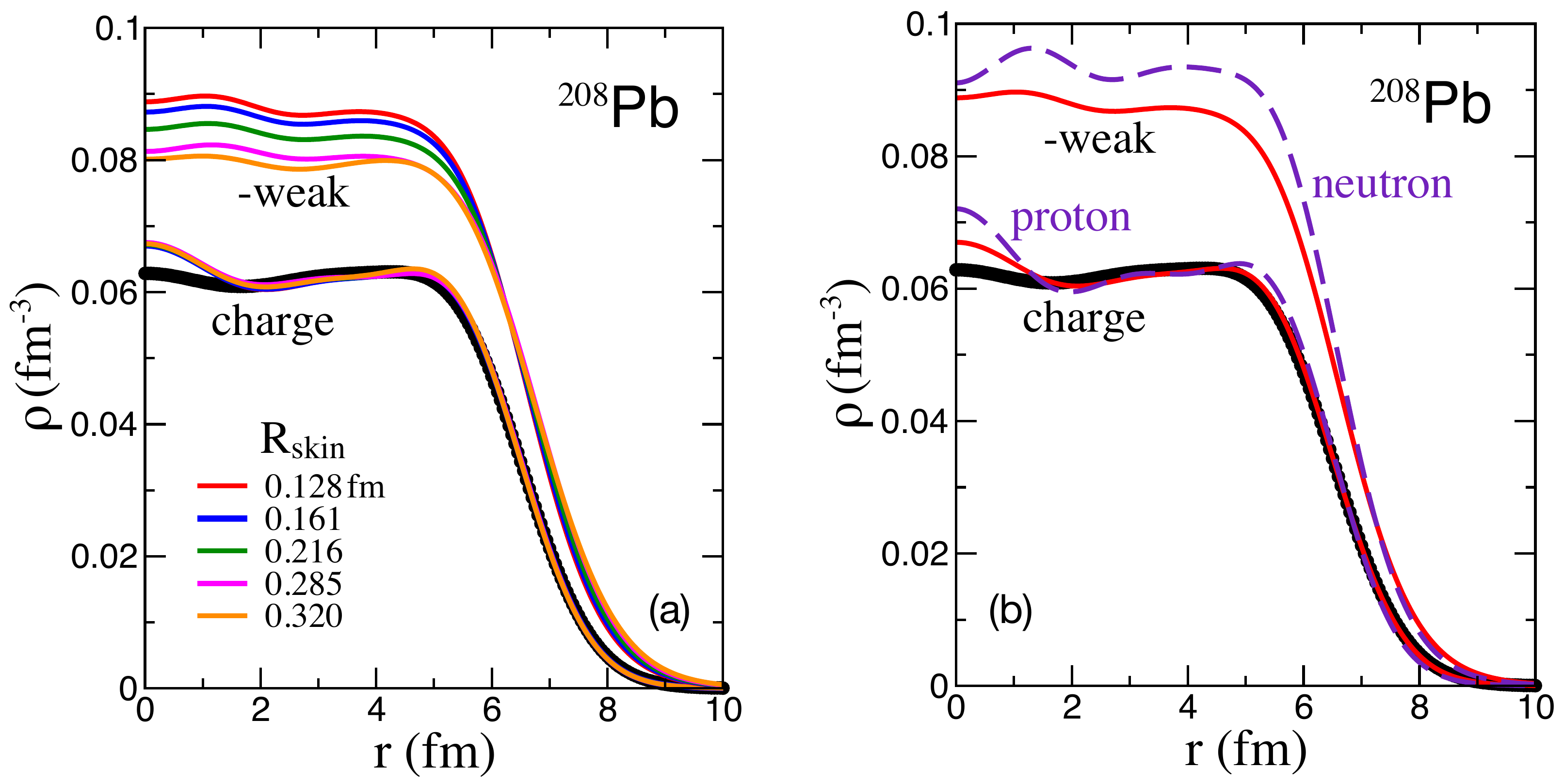}
\caption{(Color online) (a) The charge density of ${}^{208}$Pb as predicted by the set of relativistic
mean-field models introduced in the text; the experimental data (shown in black) is from 
Ref.\,\cite{DeJager:1987qc}. 
Given that all models are informed by the charge radius of ${}^{208}$Pb, there are no significant 
differences in their predictions---even in the nuclear interior. In contrast, differences in the weak charge 
density are clearly visible as a result of the calibration procedure. (b) ``Point" proton and neutron 
densities as predicted by one of the five models are compared against the corresponding charge 
and weak charge densities. The figure illustrates that whereas the charge density follows closely
the proton distribution, the weak charge density is largely driven by the neutron distribution.}
\label{Fig5}
\end{figure*}

Theoretical predictions for the charge and weak-charge densities of ${}^{208}$Pb are displayed in Fig.\ref{Fig5}(a) 
alongside the experimental charge density\,\cite{DeJager:1987qc}. Now the model dependence is evident: 
models predicting a small weak (or neutron) radius must be enhanced in the interior as they all must integrate 
to the same weak charge. In contrast, all theoretical predictions for the charge density fall within a single narrow 
band, as a natural consequence that all were fitted to the experimental charge radius of ${}^{208}$Pb. The various 
models displayed in the figure have been labeled according to the value predicted for the neutron skin thickness. 
In turn ``point" proton and neutron densities, shown by the dashed lines in Fig.\ref{Fig5}(b), as predicted by the 
model with the smallest neutron skin have been included for comparison. This comparison underscores that 
whereas the charge distribution follows closely the proton density, the weak-charge density is mostly sensitive 
to the neutron distribution.

\subsubsection{PREX and CREX}

Having introduced the various theoretical models and their impact in predicting ground state densities and form 
factors of ${}^{208}$Pb, we now turn our attention to the upcoming PREX-II and CREX campaigns at JLab. In
this context it is particularly useful to frame the discussion in terms of the ``weak-skin" form factor---defined as 
the difference between the corresponding charge and weak form factors\,\cite{Thiel:2019tkm}:
\begin{equation}
 F_{\rm Wskin}(q) \equiv F_{\rm ch}(q)-F_{\rm wk}(q), 
 \hspace{5pt} {\rm with} \hspace{3pt} q\!\equiv\!\sqrt{Q^{2}}.
\label{FWskin}
\end{equation}
Although not as physically intuitive as the neutron skin thickness, the weak-skin form factor has the virtue of being
both a strong  isovector indicator and a model independent observable. Following the the low-$q$ expansion outlined 
in Eq.\,(\ref{Moments1}), the leading behavior of the weak-skin form factor is given by
\begin{equation}
 F_{\rm Wskin}(q) \approx \frac{q^{2}}{6}\Big(R_{\rm wk}^{2} - R_{\rm ch}^{2}\Big) 
                            = \frac{q^{2}}{6}\Big(R_{\rm wk}+R_{\rm ch}\Big)\Big(R_{\rm wk}-R_{\rm ch}\Big) 
                            \equiv \frac{q^{2}}{6}\Big(R_{\rm wk}+R_{\rm ch}\Big)R_{\rm Wskin}.
\label{WKskin}
\end{equation}
Thus the leading behavior of the weak-skin form factor is determined by the ``weak skin thickness"
$R_{\rm Wskin}\!\equiv\!R_{\rm wk}\!-\!R_{\rm ch}$, a quantity that incorporates information about single-nucleon 
(both charge and weak-charge) form factors. Hence $R_{\rm Wskin}$, unlike the neutron skin thickness, is a 
genuine physical observable that may in principle be extracted from experiment. In practice, however, both PREX-II 
and CREX measurements are carried out at a single value of the momentum transfer that is sufficiently large to 
invalidate the low-$q$ expansion. Thus, some model-dependent assumptions, although mild, are required to extract 
the weak skin and ultimately the neutron skin.

\begin{figure*}[ht]
\smallskip
 \includegraphics[width=0.9\columnwidth]{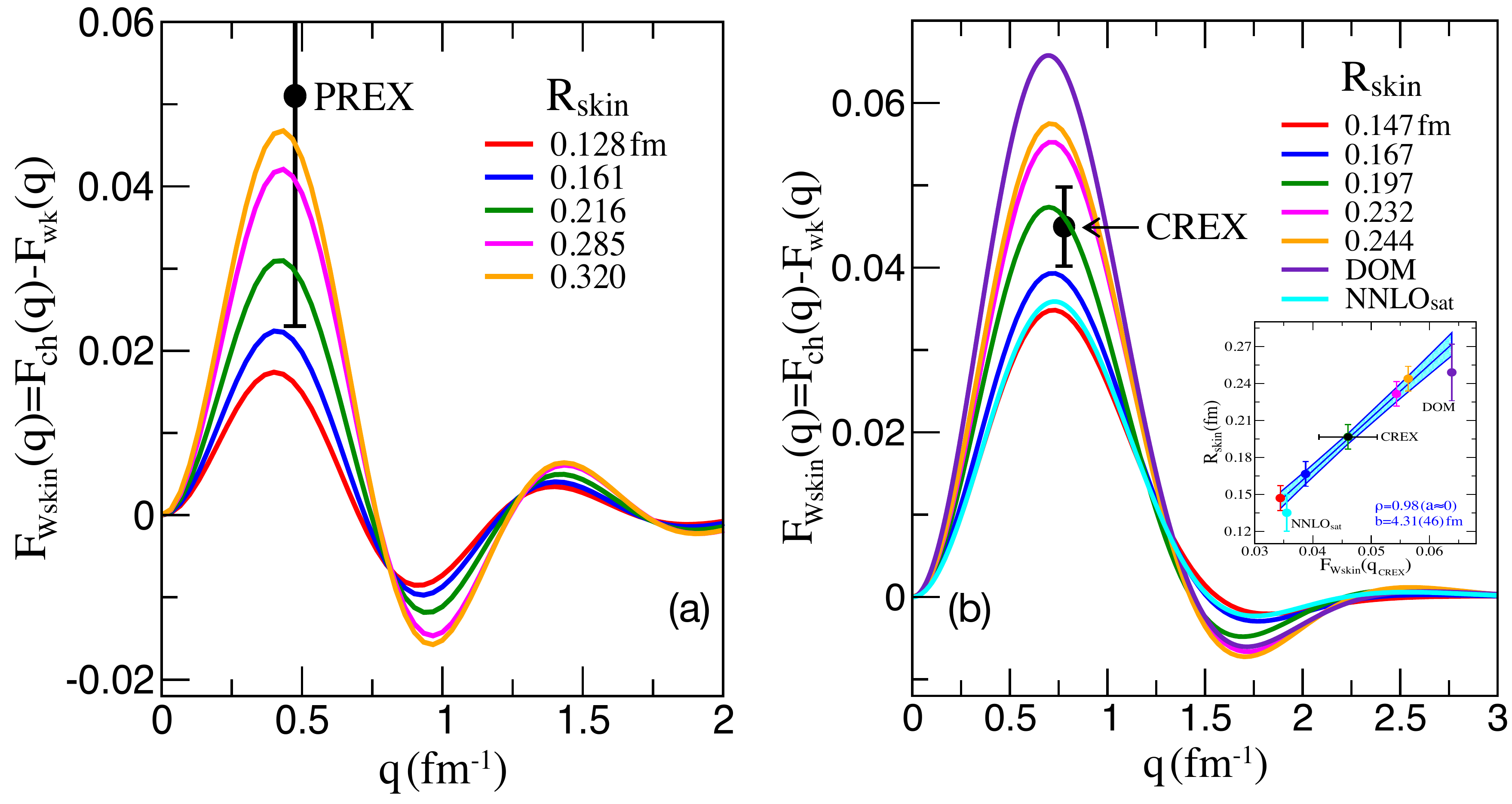}
\caption{(Color online) (a) The weak skin form factor of ${}^{208}$Pb as predicted by the set of relativistic
mean-field models introduced in the text is compared against the PREX 
measurement\,\cite{Abrahamyan:2012gp,Horowitz:2012tj}. (b) The corresponding figure for 
${}^{48}$Ca, but now including the predictions from the dispersive optical model\,\cite{Mahzoon:2017fsg} 
and the NNLO$_{\rm sat}$ interaction\,\cite{Hagen:2015yea}. The location of the CREX point is arbitrary 
but includes realistic experimental errors. Finally, the inset displays the correlation between the 
neutron skin thickness and the weak skin form form factor at the CREX momentum transfer of 
$q_{{}_{\rm CREX}}\!=\!0.778\,{\rm fm}^{-1}$.}
\label{Fig6}
\end{figure*}

To illustrate these ideas we display in Fig.\,\ref{Fig6} (borrowed from Ref.\,\cite{Thiel:2019tkm}) predictions 
for the weak-skin form factor of ${}^{208}$Pb and ${}^{48}$Ca using the same five models introduced in the
previous subsection. Also shown is the original PREX result and a tentative CREX point placed at an arbitrary 
central value but with a realistic experimental error bar\,\cite{CREX:2013}. The big model spread near the 
momentum transfer of the experiment ($q\!=\!0.475\,{\rm fm}^{-1}$ for PREX and  
$q\!=\!0.778\,{\rm fm}^{-1}$ for CREX) is easy to understand: given the model independence of the 
charge form factor, models with thicker neutron skins (or equivalently larger weak radii) predict a weak form 
factor that falls faster with momentum transfer, resulting in a larger $F_{\rm Wskin}$ at the momentum transfer 
of the experiment. In the case of ${}^{208}$Pb, we are confident that the improved PREX-II measurement
that aims to reduce the error bars by at least a factor of three will provide stringent constraints on the isovector 
sector of the nuclear density functional. Note that at the time of this writing the PREX-II campaign was already 
in full swing. Figure\,\ref{Fig6}(b) shows the corresponding plot but for the case of ${}^{48}$Ca. Together
with our theoretical predictions we also include predictions from theoretical models with a more microscopic 
underpinning. These are the dispersive optimal model (DOM) of Ref.\,\cite{Mahzoon:2017fsg} and the 
chirally-inspired NNLO$_{\rm sat}$ model of Hagen and collaborators\,\cite{Hagen:2015yea}. As is clearly
evident in the figure, the discrepancy between the two microscopic models is fairly large and spans the entire
range of mean-field models. Indeed, the values reported for the neutron skin thickness of ${}^{48}$Ca are 
$0.12\lesssim\!R_{\rm skin}^{\,48}\!\lesssim0.15\,{\rm fm}$\,\cite{Hagen:2015yea} and 
$R_{\rm skin}^{\,48}\!=\!0.249(23)\,{\rm fm}$\,\cite{Mahzoon:2017fsg}, respectively. As mentioned earlier, the
momentum transfer of the experiment is too large to justify the Taylor series expansion carried out in
Eq.\,(\ref{Moments1}). Nevertheless, as displayed in the inset, the correlation between the neutron skin thickness 
of ${}^{48}$Ca and the weak form factor at the experimental momentum transfer is strong for the limited 
set of models considered here. Values for the correlation coefficient $\rho$, slope $b$, and intercept $a\!\approx\!0$ 
are listed in the figure, while the blue region encompasses the one-sigma uncertainty band. In the case of ${}^{208}$Pb, 
a similar correlation was found using an even larger set of models; see Fig.\,3 in Ref.\,\cite{Furnstahl:2001un}.

\subsubsection{Coherent elastic neutrino nucleus scattering}

Given the imminent start of both the PREX-II and CREX campaigns at JLab, it is timely to explore the insights 
that one may gain from such compelling experiments. Specifically, we want to explore the limits that PREX-II 
and CREX may impose on the weak form factor of ${}^{132}$Xe and ${}^{40}$Ar at low momentum transfer. 
Both liquid noble gases, xenon and argon are currently being used as active targets for the detection of  
neutrinos as well as dark matter particles. In the context of \Cevens\hspace{-3pt}, the coherent cross section 
at forward angles is dominated by the weak charge of the nucleus---a quantity that may provide a portal to new 
physics due to its dependence on $\sin^{2}\theta_{\rm W}$. Away from $Q^{2}\!=\!0$, but still at low momentum
transfers, nuclear-structure corrections to the coherent cross section are dominated by the weak radius. This 
``loss of coherence", namely, the weakening of the cross section due to the deviations from the \emph{quadratic} 
enhancement with the number of target neutrons ($Q_{\rm wk}^{2}\!\approx\!N^{2}$) is a natural consequence 
of the finite nuclear size. As such, \Cevens also provides fundamental information on the neutron distribution. 

\begin{figure*}[ht]
\smallskip
 \includegraphics[width=0.9\columnwidth]{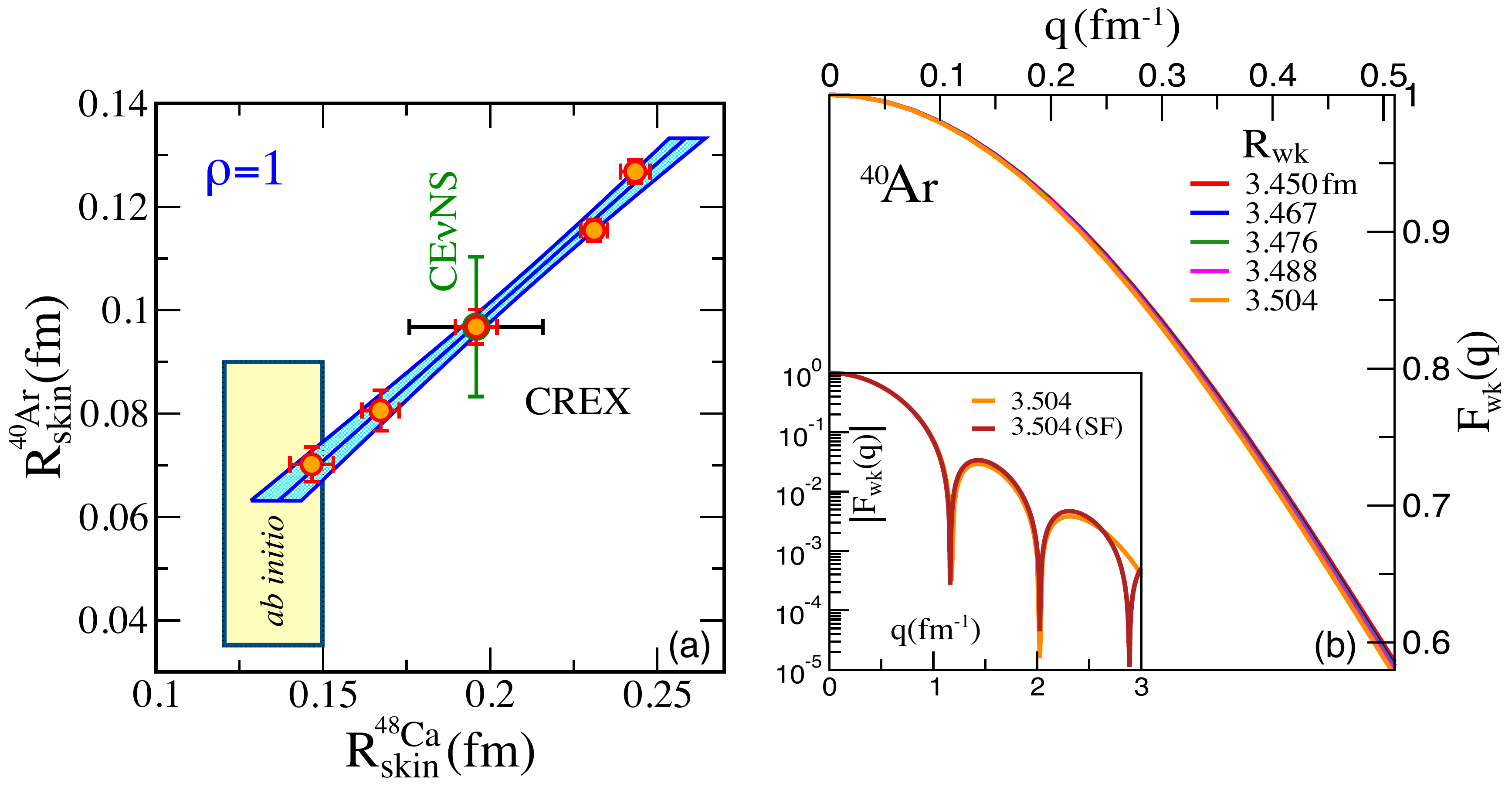}
 \caption{(Color online) (a) Data-to-data relation between the neutron skin thickness of ${}^{48}$Ca and 
the correspondent skin thickness of ${}^{40}$Ar. Theoretical error bars are displayed for each model 
together with error bands for the optimal linear fit: 
$R_{\rm skin}^{\,40}\!=\!-0.015\!+\!0.572\,R_{\rm skin}^{\,48}$. Note that whereas the CREX error 
bar is realistic, the central value is arbitrarily placed at $R_{\rm skin}^{\,48}\!\approx\!0.2\,{\rm fm}$. 
The rectangular section includes the predictions from Refs.\,\cite{Hagen:2015yea,Payne:2019}. 
(b) The weak form factor of ${}^{40}$Ar as predicted by the set of relativistic mean-field models introduced
in the text. Given that the form factor at low momentum transfers is sensitive to the weak radius---and
not to the weak skin---differences in the model predictions are very small. The inset shows a comparison 
between one of the self-consistent mean field models and the corresponding symmetrized Fermi function.} 
\label{Fig7}
\end{figure*}

Possible constraints on the neutron skin thickness of ${}^{40}\!{\rm Ar}$ deduced from the upcoming CREX measurement 
are depicted in Fig.\,\ref{Fig7}(a). Based on the particular set of energy density functionals used in this work, we found an 
extremely strong correlation ($\rho\!\approx\!1$) between the neutron skin thickness of ${}^{48}$Ca and ${}^{40}$Ar. Theoretical 
errors were computed from the corresponding covariance matrix extracted from the calibration procedure\,\cite{Chen:2014sca}. 
In turn, the optimal straight line and the associated error band depicted in the figure were obtained from a linear regression 
analysis.  Assuming, as in Fig.\,\ref{Fig6}(b), a central CREX value placed \emph{arbitrarily} at 
$R_{\rm skin}^{\,48}\!\approx\!0.2\,{\rm fm}$---but accurately reflecting the anticipated experimental error---results in a 
neutron skin thickness for argon of $R_{\rm skin}^{\,40}\!=\!0.097(14)\,{\rm fm}$. The rectangular section attached to the 
figure reflects predictions from microscopic models of the neutron skin thickness of ${}^{48}$Ca\,\cite{Hagen:2015yea} 
and ${}^{40}$Ar\,\cite{Payne:2019}. These predictions suggest a fairly soft symmetry energy; indeed, the predicted value 
for the slope of the symmetry energy falls in the fairly narrow range of
$L\!=\!37.8$\,--\,$47.7$\,MeV\,\cite{Hagen:2015yea}. The associated weak form factor of ${}^{40}$Ar 
is displayed in Fig.\ref{Fig7}(b) as a function of momentum transfer. Recall that for spinless nuclei, the weak form factor 
encodes the entire nuclear-structure contribution to the coherent cross section. The momentum-transfer range extends up 
to a maximum value of $q_{\rm max}\!=\!0.5\,{\rm fm}^{-1}$ which, in turn, corresponds to a maximum incoming 
neutrino energy of $E_{\rm max}\!\simeq\!q_{\rm max}/2\!\approx\!50\,{\rm MeV}$ and a maximum recoil energy of 
$T_{\rm max}\!\approx\!130\,{\rm keV}$. 
Although the weak form factor falls off to almost half of its maximum value, resulting 
in a significant quenching of the coherent cross section from its $Q^{2}\!=\!0$ value, differences in the model predictions 
amount to less than  2\% over the entire---yet relatively modest---momentum-transfer range. This mild model dependence 
stands in stark contrast to the large spread displayed by the neutron skin thickness, an isovector quantity that emerges as 
the small difference of two large numbers. The insensitivity of the coherent cross section to nuclear-structure effects---especially 
for nuclei with a small neutron excess---makes \Cevens an ideal probe of new physics\,\cite{UNAM:2019}. Although compelling,
it is essential to validate this claim against a set of theoretical models that rely on a different protocol. Finally, we display in 
the inset of Fig.\ref{Fig7}(b) a comparison between a symmetrized Fermi function\,\cite{Sprung:1997,Piekarewicz:2016vbn} 
informed by the ``stiffest" density functional. Encapsulated in the shape of the form factor are the characteristic diffractive 
oscillations driven by the half-density radius $c$ and the exponential falloff controlled by the diffuseness parameter $a$; 
see Eq.\,(\ref{HighqFs}). By adjusting $c$ and $a$ to reproduce the first two moments of the selected density functional, 
good agreement is obtained for momentum transfers $q\!\gtrsim\!2\,{\rm fm}^{-1}$.

\begin{figure*}[ht]
\smallskip
 \includegraphics[width=0.9\columnwidth]{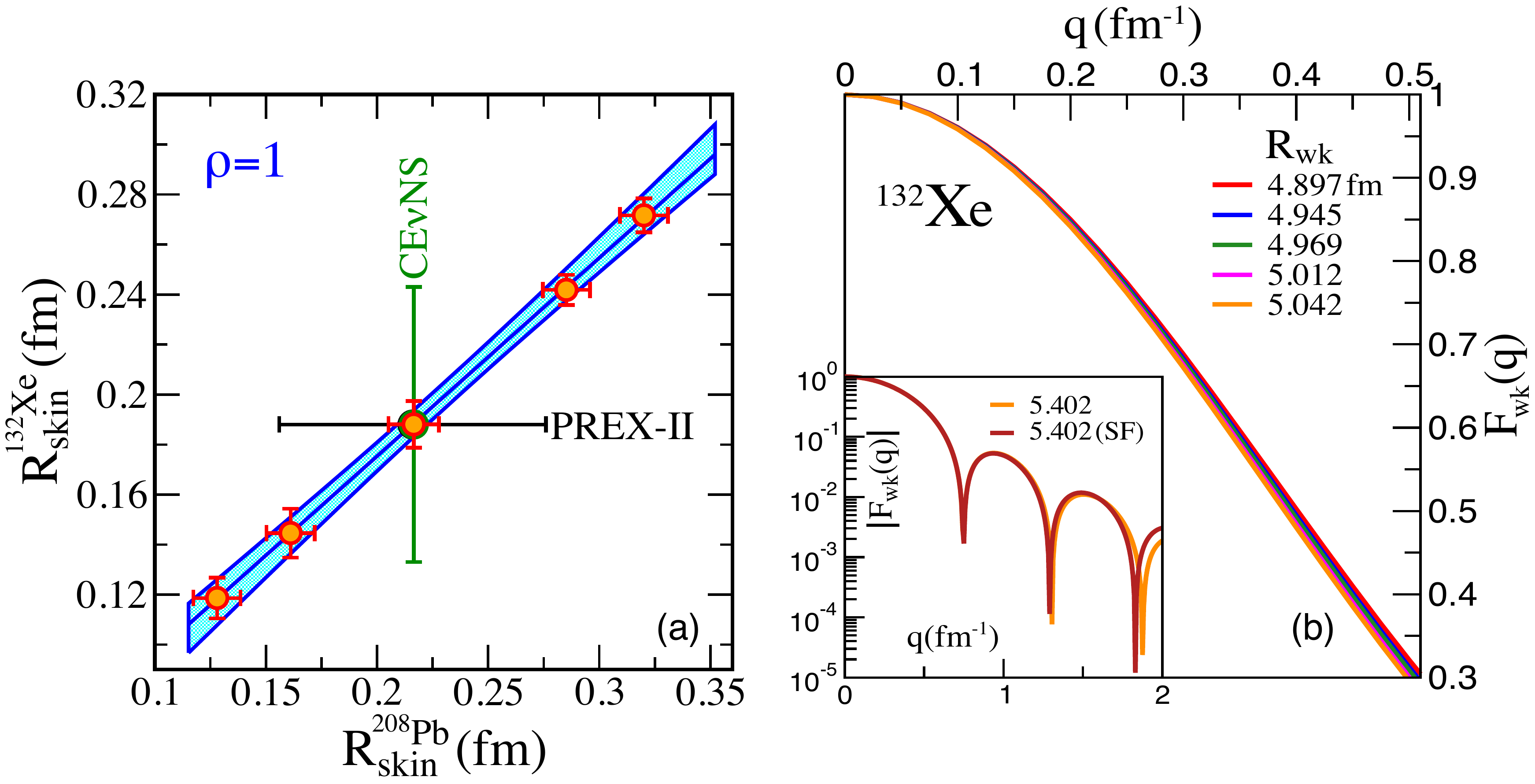}
 \caption{(Color online) (a) Data-to-data relation between the neutron skin thickness of ${}^{208}$Pb and 
the correspondent skin thickness of ${}^{132}$Xe. Theoretical error bars are displayed for each model 
together with error bands for the optimal linear fit: 
$R_{\rm skin}^{\rm 132}\!=\!0.017\!+\!0.793\,R_{\rm skin}^{\rm 208}$. Note that whereas the PREX-II error 
bar is realistic, the central value is arbitrarily placed at $R_{\rm skin}^{132}\!\approx\!0.22\,{\rm fm}$. (b) 
The weak form factor of ${}^{132}$Xe as predicted by the set of relativistic mean-field models introduced
in the text. Given that the form factor at low momentum transfers is sensitive to the weak radius---and
not to the weak skin---differences in the model predictions are very small. The inset shows a comparison 
between one of the self-consistent mean field models and the corresponding symmetrized Fermi function.} 
\label{Fig8}
\end{figure*}

The analogous plot, but now displaying the correlation between ${}^{208}$Pb and ${}^{132}$Xe, is shown in
Fig.\,\ref{Fig8}. The PREX collaboration main goal is to determine the neutron skin thickness of ${}^{208}$Pb
in order to constrain the density dependence of the symmetry energy and ultimately the structure of neutron 
stars. Qualitatively, ${}^{208}$Pb is as efficient in constraining the neutron skin thickness of ${}^{132}$Xe as 
${}^{48}$Ca is in constraining the corresponding skin thickness of ${}^{40}$Ar. This result could have been 
anticipated given the strong correlation found in Ref.\,\cite{Piekarewicz:2012pp} between the neutron skin 
thickness of ${}^{208}$Pb and ${}^{132}$Sn, where a much larger set of both relativistic and nonrelativistic 
density functionals was used to validate the correlation. Quantitatively, however, the anticipated 
$\sim\!0.06\,{\rm fm}$ from the PREX-II measurement translates into a larger error for ${}^{132}$Xe than 
for ${}^{40}$Ar. That is, assuming a central PREX-II value \emph{arbitrarily} placed at
 $R_{\rm skin}^{208}\!\approx\!0.22\,{\rm fm}$ yields a neutron skin thickness for xenon of 
 $R_{\rm skin}^{132}\!=\!0.188(55)\,{\rm fm}$. Yet, in the case of the weak form factor of ${}^{132}$Xe
displayed in Fig.\,\ref{Fig8}(b), the weak model dependence found in ${}^{40}$Ar still persists---a fact that 
reinforces \Cevens as a powerful tool for the search for new physics\,\cite{AristizabalSierra:2019zmy}. 
In addition, we continue to find excellent agreement between the (analytic) two-parameter symmetrized 
Fermi function and the numerically generated weak form factor over a significant momentum-transfer range.

We conclude by underscoring the solid underpinning of the mild model dependence of the weak form 
factor. Within the scope of density functional theory, the calibration of the functional is informed by both 
binding energies and charge radii of a variety of spherical nuclei. Hence, although some flexibility remains
in the determination of the neutron (or weak) skin thickness because uncertainties in the isovector sector, 
this flexibility is not without limits. Indeed, the quality of the fit deteriorates considerably once an overly
thin or thick neutron skin in ${}^{208}$Pb is assumed. In other words, the distribution of electric charge 
informs, to some extent, the corresponding distribution of weak charge. Nevertheless,  given the enormous 
precision demanded to uncover new physics, it becomes imperative to minimize the theoretical uncertainties 
in the determination of the weak form factor\,\cite{AristizabalSierra:2019zmy}. Both CREX and PREX-II will 
be instrumental in realizing this goal.

\subsubsection{Charge radii of mirror nuclei}

We conclude this section by examining correlations between the neutron skin thickness of 
${}^{48}$Ca and the difference in proton radii of mirror nuclei. The underpinning of the correlation is 
compelling: in the limit of exact charge symmetry, the neutron skin of a given nucleus is identical to
the difference in proton radii between the nucleus of interest and its mirror nucleus\,\cite{Brown:2017}. 
Indeed, in the absence of charge-symmetry violations the \emph{entire} neutron density of the given
nucleus must be identical to the proton density of its mirror nucleus. Given the long and successful 
history of electron scattering experiments in mapping the proton distribution, perhaps this connection 
can help constrain the neutron distribution.  Naturally, charge symmetry is broken at the most 
fundamental level by quark-mass differences and electromagnetic effects\,\cite{Miller:2006tv}. As 
we have done in our earlier work\,\cite{Yang:2017vih}, we limit ourselves here to study charge-symmetry 
violations induced by the Coulomb interaction, which although critical, is well understood. Not included in 
our description, however, are explicit manifestations of charge symmetry violations in the nuclear 
force---known to be responsible for residual differences in the binding energy of mirror nuclei, 
the so-called Okamoto-Nolen-Schiffer anomaly\,\cite{Okamoto:1964pc,Nolen:1969ms}.

\begin{figure*}[ht]
\smallskip
 \includegraphics[width=0.9\columnwidth]{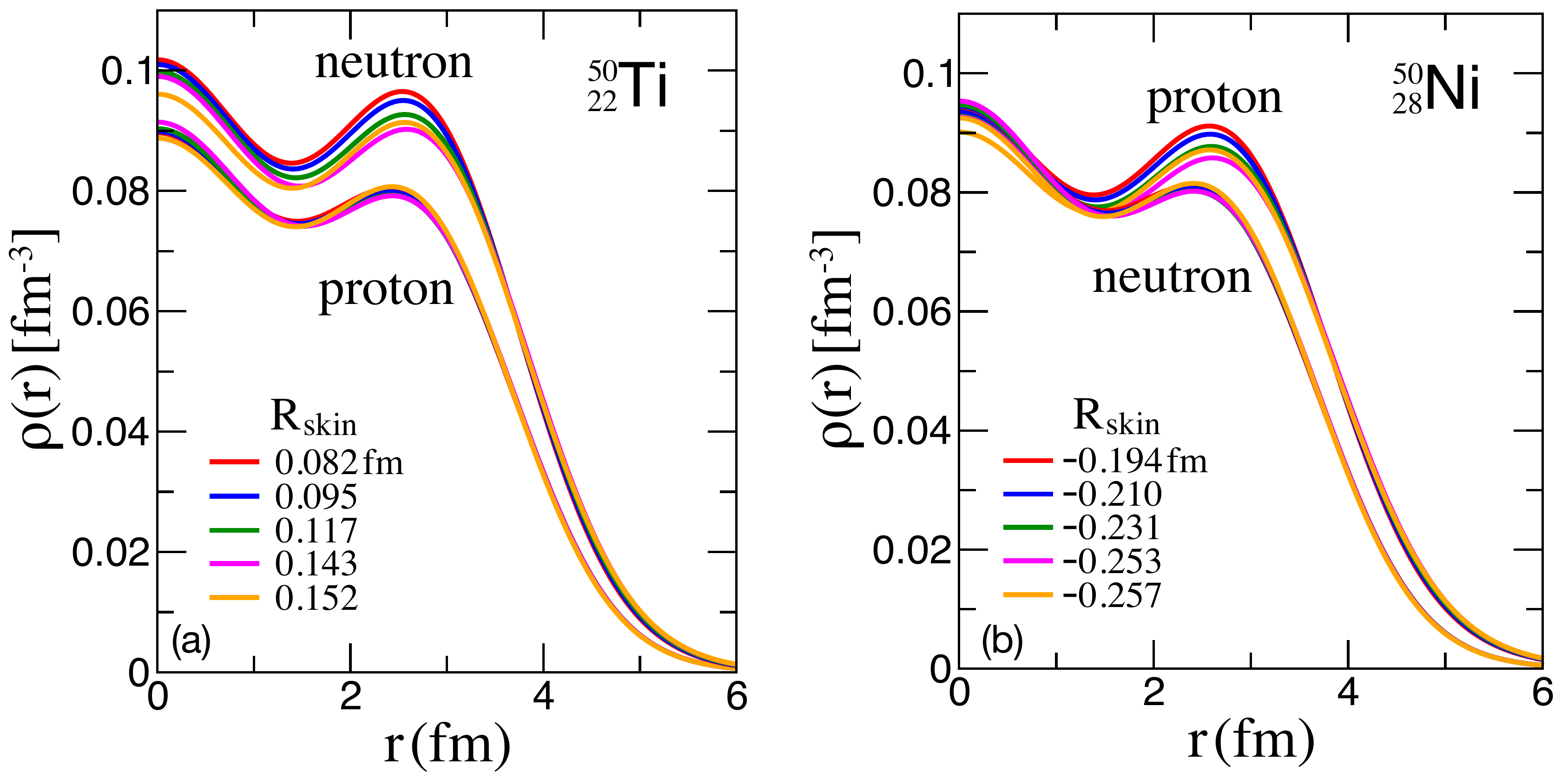}
\caption{(Color online) (a) Point proton and neutron densities for ${}^{50}$Ti as predicted by the set 
of relativistic mean-field models introduced in the text. Given that the charge radius of neighboring 
${}^{48}$Ca was included in the calibration of the models, we see a small spread in the predictions 
of the proton density. However, since these models include different predictions for the neutron radius 
of ${}^{208}$Pb, the neutron distribution of ${}^{50}$Ti shows a significantly larger spread. (b) In
contrast, the corresponding plot for the neutron deficient ${}^{50}$Ni nucleus displays the opposite 
trend, namely, a large spread in the proton density and a small spread in the neutron distribution.}
\label{Fig9}
\end{figure*}

Ideally, one would like to explore correlations between the two doubly-magic mirror nuclei ${}^{48}$Ca 
and ${}^{48}$Ni. However, given that ${}^{48}$Ni is unstable against two-proton decay, we concentrate 
instead on the neighboring ${}^{50}$Ti-${}^{50}$Ni mirror pair. In Fig.\,\ref{Fig9}(a) we display proton and 
neutron densities for ${}^{50}$Ti as predicted by the same five relativistic density functionals employed 
throughout this work; Coulomb effects are fully incorporated into these calculations. Having six excess 
neutrons, ${}^{50}$Ti develops a neutron skin that is slightly smaller than ${}^{48}$Ca. Even so, the 
theoretically-induced spread in the neutron density is clearly discernible. The situation is drastically 
different for the proton density, which being informed by the charge radius of neighboring ${}^{48}$Ca, 
displays a modest model dependence. Indeed, charge (or proton) radii differ from each other by at 
most 0.3\% and by less than 2\% from the experimental value\,\cite{Angeli:2013}.
As anticipated, Fig.\,\ref{Fig9}(b) is---at least qualitatively---the ``mirror" image of Fig.\,\ref{Fig9}(a). Now 
the theoretical spread in the neutron density is barely noticeable while the large model spread has shifted 
to the proton density. Interestingly, the size of the ``proton skin" in ${}^{50}$Ni is significantly larger than 
the corresponding neutron skin in ${}^{50}$Ti. This is because in the neutron deficient ${}^{50}$Ni isotope 
both the Coulomb repulsion and the symmetry energy work in tandem in pushing protons out to the surface, 
thereby creating a larger proton skin.

\begin{figure*}[ht]
\smallskip
 \includegraphics[width=0.9\columnwidth]{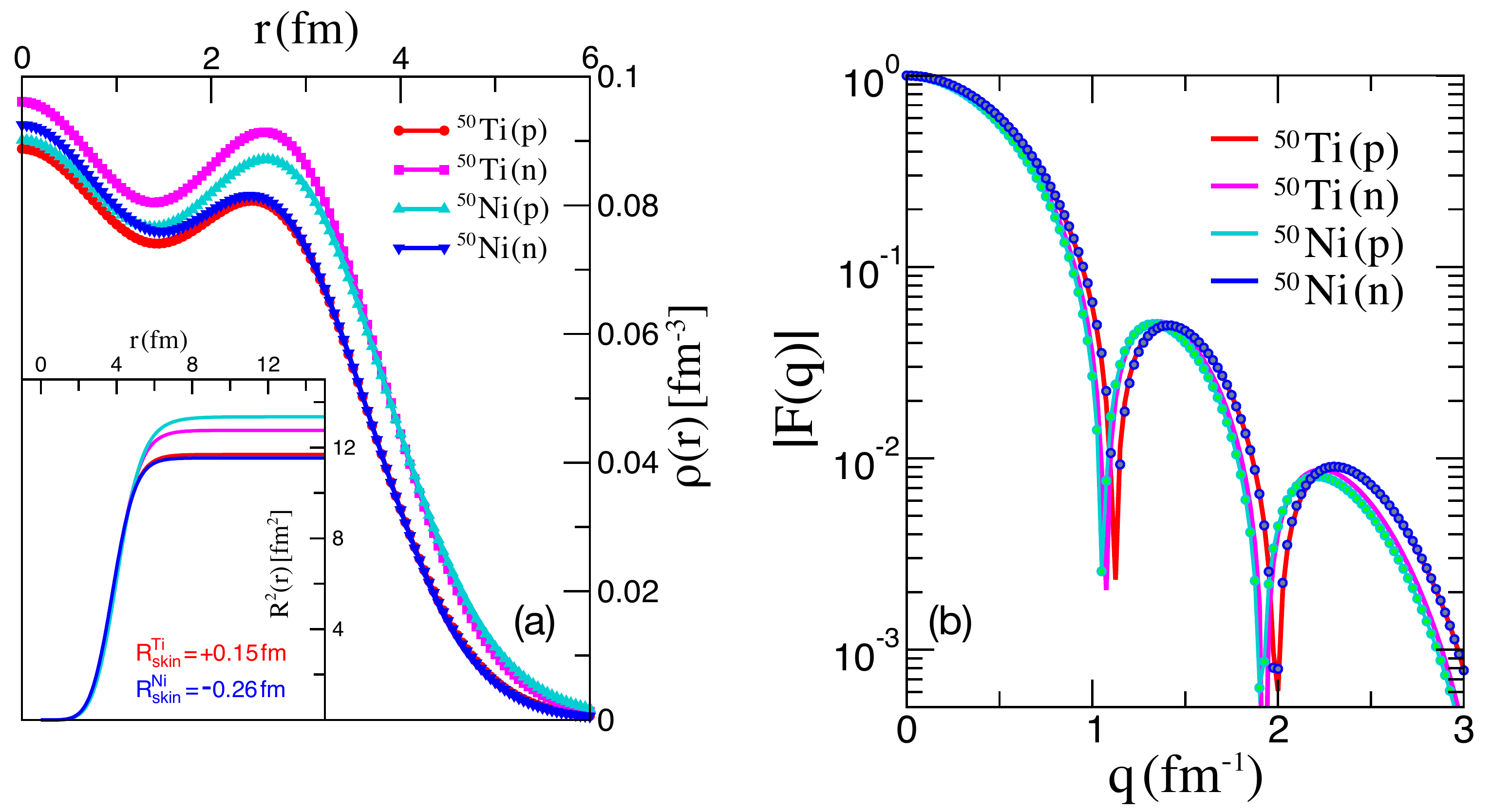}
\caption{(Color online) (a) Point proton and neutron densities for both ${}^{50}$Ti and ${}^{50}$Ni 
using (for simplicity) only one of the theoretical models introduced in the text. Shown in the inset is 
the ``running sum" defined in Eq.\,(\ref{R2r}). Note that the mean square radii of the minority species 
(i.e., protons for ${}^{50}$Ti and neutrons for ${}^{50}$Ni) are practically indistinguishable. (b) The 
corresponding plot for the point proton and neutron form factors. The figure underscores the excellent
agreement between the form factors, especially for the case of the minority species, i.e., protons in 
titanium and neutrons in nickel.}
\label{Fig10}
\end{figure*}

To further strengthen the connection between mirror nuclei, we display together in Fig.\,\ref{Fig10}(a) proton and 
neutron densities for both ${}^{50}$Ti and ${}^{50}$Ni; for clarity we employ only one of the models. Although the 
matching appears far from ideal, most of the differences are confined to the nuclear interior which is known to be 
sensitive to the behavior of the associated form factor at high momentum transfers. Instead, the agreement in the 
tails of the spatial distribution---associated to the low momentum behavior of the form factor---is significantly better. 
To fully appreciate this point, we display in the inset the ``running sum" associated to the mean square radius. Taking 
the charge radius as an example, the running sum is defined as:
\begin{equation}
 R_{\rm ch}^{2}(r) = \frac{1}{Z} \int_{0}^{r} 4\pi\,x^{4} \rhoX{\rm ch}(x)dx
 \hspace{3pt}\xrightarrow{r\rightarrow\infty} \hspace{3pt} R^{2}_{\rm ch}.
 \label{R2r}
\end{equation}
There is a fairly good agreement between the root-mean-square neutron radius in titanium ($R_{n}\!=\!3.572\,{\rm fm}$) 
and the corresponding proton radius in nickel ($R_{p}\!=\!3.654\,{\rm fm}$). In both cases the symmetry pressure pushes
the ``majority'' species (protons in nickel and neutrons in titanium) to the surface. However, in the case of nickel the
additional Coulomb repulsion pushes the protons even further out to the surface. Whereas such a modest $\sim\!2\%$ 
disagreement may have been expected, the minute 0.6\% discrepancy in the radii of the ``minority" species is better 
than anticipated: $R_{p}\!=\!3.419\,{\rm fm}$ for titanium against $R_{n}\!=\!3.397\,{\rm fm}$ for nickel. One would have 
expected that because of the Coulomb repulsion, the 22 protons in titanium would be pushed farther out than the 22 
neutrons in nickel; and they do---but only by a mere $\sim\!0.02\,{\rm fm}$. That the neutron distribution in nickel is more 
extended than anticipated may be due to their tendency to track the 28 protons. More remarkable, is that this level 
of  agreement extends to the form factors over a significant range of momentum transfers;  see Fig.\,\ref{Fig10}(b). 
Although in the interest of clarity our results are displayed for only one model, the same level of agreement is observed 
for all the other models considered in this work. Indeed, so far we have found this behavior to be 
\emph{universal}---independent of both the model and the mirror partners under consideration. And while we believe 
that there are compelling theoretical arguments underpinning this behavior, it would be valuable for our results to be 
tested against alternative approaches.

\begin{figure*}[ht]
\smallskip
 \includegraphics[width=0.9\columnwidth]{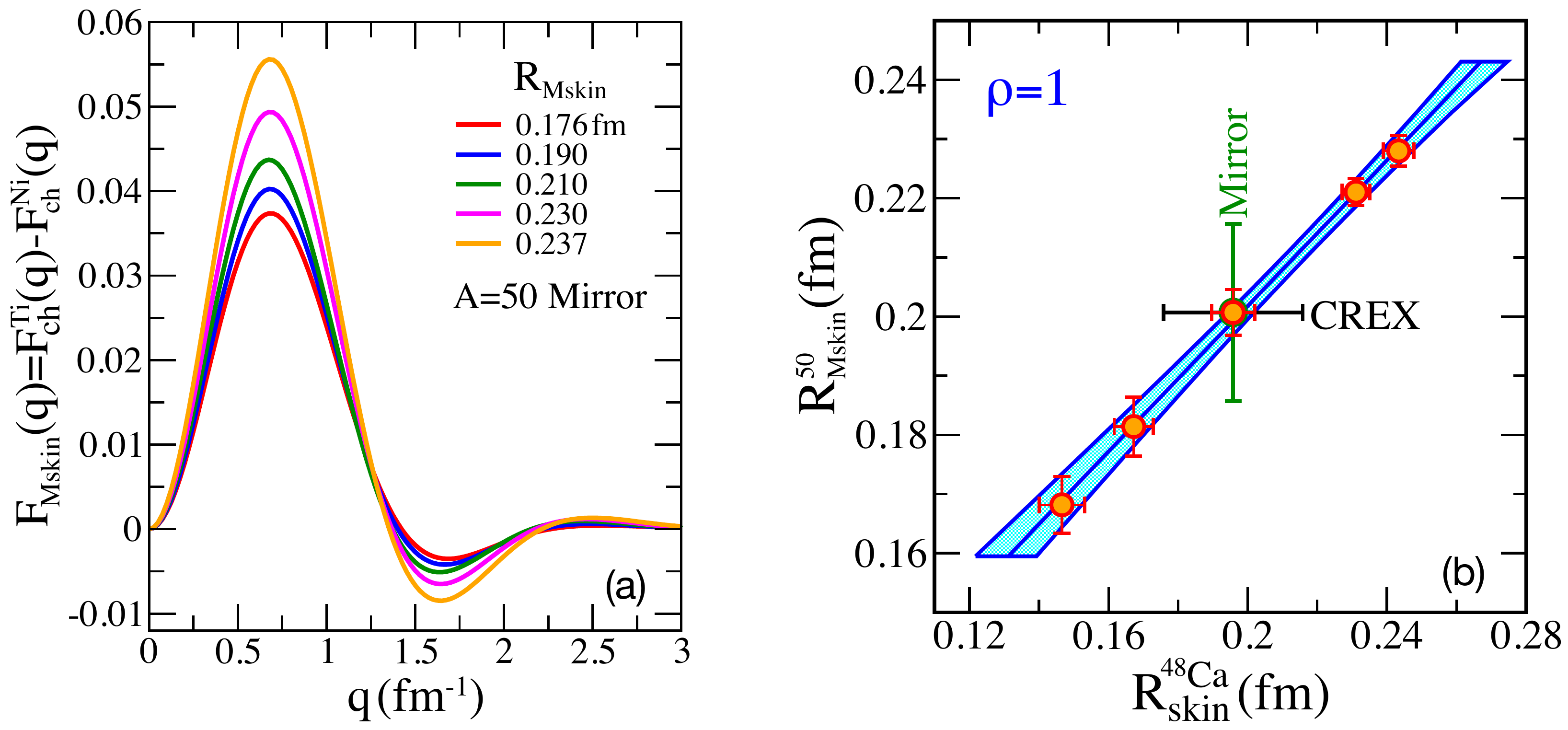}
\caption{(Color online) (a) The ``mirror skin" form factor of the $A\!=\!50$ mirror nuclei titanium and nickel;
this plot should be compared against the corresponding weak skin form factor of ${}^{48}$Ca displayed 
in Fig.\,\ref{Fig6}(b). (b) Data-to-data relation between the neutron skin thickness of ${}^{48}$Ca and 
the mirror skin thickness of the $A\!=\!50$ mirror nuclei. Theoretical error bars are displayed for each 
model together with error bands for the optimal linear fit: 
$R_{\rm Mskin}^{\rm 50}\!=\!0.079\!+\!0.615\,R_{\rm skin}^{\,48}$. Note that whereas the CREX error 
bar is realistic, the central value is arbitrarily placed at $R_{\rm skin}^{\,48}\!\approx\!0.2\,{\rm fm}$.} 
\label{Fig11}
\end{figure*}

Having established the close connection between the ground-state form factors of mirror nuclei---even in the 
presence of Coulomb interactions---we now define the ``mirror skin" form factor in complete analogy to the 
weak skin form factor introduced in Eq.\,(\ref{FWskin}). That is, 
\begin{eqnarray}
 F_{\rm Mskin}(Z,N;q) \equiv F_{\rm ch}(Z,N;q)-F_{\rm ch}(N,Z;q) & \approx & 
                                     \frac{q^{2}}{6}\Big(R_{\rm ch}^{2}(N,Z) - R_{\rm ch}^{2}(Z,N)\Big) \nonumber\\
                            &=&  \frac{q^{2}}{6}\Big(R_{\rm ch}(N,Z)+R_{\rm ch}(Z,N)\Big)
                                                           \Big(R_{\rm ch}(N,Z)-R_{\rm ch}(Z,N)\Big) \nonumber\\
                      &\equiv&  \frac{q^{2}}{6}\Big(R_{\rm ch}(N,Z)+R_{\rm ch}(Z,N)\Big)R_{\rm Mskin}(Z,N),
\label{FMskin}
\end{eqnarray}
where we have defined the ``mirror skin thickness" as 
$R_{\rm Mskin}(Z,N)\!\equiv\!R_{\rm ch}(N,Z)\!-\!R_{\rm ch}(Z,N)$.

The mirror skin form factor for the $A\!=\!50$ (Ti-Ni) mirror pair is displayed in Fig.\,\ref{Fig11}(a). The uncanny 
resemblance between this figure and Fig.\,\ref{Fig6}(b) for the weak skin form factor of ${}^{48}{\rm Ca}$ is no 
accident and serves to reinforce the notion that the mirror skin serves as a reliable proxy for the weak 
skin\,\cite{Brown:2017,Yang:2017vih,Sammarruca:2017siq}. Moreover, the resemblance between the two
figures indicates that this intimate connection extends beyond $Q^{2}\!=\!0$ (i.e., skins) to the entire form
factor. To conclude---and to further reinforce the connection---we display in Fig.\,\ref{Fig11}(b) the correlation 
between mirror skins and neutron skins. As we have done in Fig.\,\ref{Fig7}(a) for the case of ${}^{40}$Ar, 
we show the error that could be inferred in $R_{\rm Mskin}^{50}$ from the upcoming CREX measurement.
Assuming a central CREX value at $R_{\rm skin}^{\,48}\!\approx\!0.2\,{\rm fm}$, we deduce a corresponding 
value for the difference in charge radii of mirror nuclei of  $R_{\rm Mskin}^{50}\!=\!0.201(15)\,{\rm fm}$. 
This suggests that a measurement of the charge radius of ${}^{50}$Ni to $0.01\,{\rm fm}$ could provide 
important theoretical constraints.

We close this section with a provocative question: could one provide a theoretical estimate for $R_{\rm Mskin}^{50}$? 
In the case of ${}^{50}$Ti, an experimental determination of its charge radius already exists\,\cite{Angeli:2013}. 
In contrast, ${}^{58}$Ni---eight neutrons away from ${}^{50}$Ni---is the most neutron-deficient nickel isotope with 
a well measured charge radius.  At present, the only experimental alternative to measure the internal structure of 
short-lived radioactive nuclei is the novel SCRIT facility at RIKEN\,\cite{Suda:2009zz}. Theoretically, however, one
may try to overcome the intrinsic limitations of existing theoretical descriptions by using machine learning. Indeed, 
an approach based on the construction of a ``Bayesian Neural Network" (BNN) has been successfully implemented 
in the refinement of nuclear masses\,\cite{Utama:2015hva,Utama:2017wqe,Utama:2017ytc,Neufcourt:2018syo,
Neufcourt:2019qvd} and charge radii\,\cite{Utama:2016rad}. The BNN paradigm is implemented in two steps.
First, one starts with an accurate model that provides a good description of the desired observable. Then, one 
refines the model by training an artificial neural network on the ``small" residuals between the theoretical predictions 
and the experimental data. Note that besides improved predictions, the Bayesian nature of the approach provides
reliable estimates of the theoretical uncertainties. In the near future one could revisit the BNN refinement of charge 
radii introduced in Ref.\,\cite{Utama:2016rad} to examine whether reliable predictions could be made for the charge 
radius of ${}^{50}$Ni---in spite of the large extrapolation that this entails.

\section{Conclusions}
\label{sec:Conclusions}

Where do the neutrons go? As recently articulated in Ref.\,\cite{Piekarewicz:2019ahf}, the elusive answer to such a 
deceptively simple question provides fundamental new insights into the structure of both atomic nuclei and neutron 
stars. Although vast experimental resources have been devoted for decades to measure the neutron density of
atomic nuclei, a model-independent determination remains elusive\,\cite{Thiel:2019tkm}. Although valuable, these
experiments involve hadronic reactions that are hindered by major uncertainties and uncontrolled approximations. 
Yet, with recent experimental developments the possibility of measuring the neutron distribution using exclusively 
electroweak probes has become a reality. In this contribution we have examined the information content of the 
following three electroweak experiments: (a) parity violating elastic electron 
scattering\,\cite{Abrahamyan:2012gp,Horowitz:2012tj}, (b) coherent elastic neutrino nucleus 
scattering\,\cite{Akimov:2017ade}, and elastic electron scattering from unstable nuclei---particularly of the 
neutron-deficient member of a mirror pair\,\cite{Suda:2009zz}. 
None of these enormously challenging experimental techniques are likely to provide a complete picture of the 
neutron distribution; although parity violating elastic electron scattering has evolved into a mature field, it is 
unrealistic to expect that it will determine the weak form factor at several values of the momentum transfer. 
In the case of \Cevens and elastic electron scattering from exotic nuclei, both experimental programs are still 
in their infancy---despite having already reached impressive milestones\,\cite{Suda:2009zz,Akimov:2017ade}. 
Yet the immense value of the various electroweak experiments is that they provide model-independent information 
that could both inform theoretical models and provide powerful anchors to guide future experiments with hadronic 
probes at rare isotope facilities. 

Following Ref.\,\cite{Thiel:2019tkm} and motivated by the ongoing PREX-II and imminent CREX campaigns, we 
affirmed the value of the ``weak skin form factor" as a model-independent observable that provides 
powerful constraints on both the neutron distribution and ultimately on the density dependence of the symmetry 
energy. Assuming realistic experimental errors, we then examined the impact of PREX-II and CREX on CE$\nu$NS.
For the set of accurately-calibrated models employed in this work, we observed a very strong correlation between 
the neutron skins of ${}^{208}$Pb and ${}^{132}$Xe; an equally strong correlation was seen between ${}^{48}$Ca 
and ${}^{40}$Ar. Both xenon and argon are liquid noble gases currently used for the detection of neutrinos and 
dark-matter particles. In particular, \Cevens provides an irreducible background to dark-matter searches, therefore
documenting nuclear-structure uncertainties is of utmost importance. Besides defining the ``neutrino floor" 
in the direct searches for dark-matter particles, \Cevens may also provide a portal to new physics. Indeed, the
coherent cross section is proportional to the square of the product of the weak nuclear charge with the weak 
nuclear form factor. Imposing stringent constraints on the weak form factor at low momentum 
transfers\,\cite{AristizabalSierra:2019zmy} may help isolate 
$\sin^{2}\!\theta_{\rm W}$\,\cite{UNAM:2019}. Although the theoretical models explored in this work are flexible 
enough to allow tuning of the (yet undetermined) neutron skin thickness of ${}^{208}$Pb, the flexibility is not 
without limits---primarily because the calibration of the models is informed by the binding energies and charge 
radii of a variety of nuclei. As a result, we found a very small spread in the predictions of the weak form factor 
at low momentum transfers.  Whether the spread is small enough to provide meaningful constraints on 
$\sin^{2}\!\theta_{\rm W}$ is currently under investigation.

Finally, we examined correlations between the neutron skin thickness of ${}^{48}$Ca and differences in the 
charge (or proton) radii of the $A\!=\!50$ mirror nuclei titanium and nickel---such a correlation emerges in the 
limit of exact charge symmetry. Indeed, in this limit the neutron radius of ${}^{48}$Ca would be identical to 
the proton radius of ${}^{48}$Ni. As has been documented earlier\,\cite{Brown:2017,Yang:2017vih,Sammarruca:2017siq}, 
we observed that the strong correlation remains valid even after restoring Coulomb effects. Note, however, 
that we have not tested the impact of explicit charge symmetry violations in the nuclear interaction, which
could become important given that the difference in charge radii emerges from the cancellation of two ``large" 
numbers. Of course, exact charge symmetry is not restricted to nuclear radii, but extends to the entire 
distributions. In this sense, the neutron distribution of a given nucleus should be identical to the proton 
distribution of its mirror partner. Surprisingly, we found that the agreement between the form factors of the 
minority species---protons in titanium and neutrons in nickel---was significantly better than anticipated,
even after restoring the Coulomb interaction. Although not discussed explicitly here, the connection among 
the minority species seems robust, as it was also observed in other mirror pairs. It would be interesting to 
test whether this robust connection also holds under other theoretical paradigms.

In summary, remarkable new experimental advances are starting to provide new insights into the neutron distribution 
using solely electroweak probes.  A model independent determination of neutron densities and their associated form 
factors has far reaching implications. For example, a highly accurate determination of the weak-charge form 
factor---which is closely related to the neutron form factor---provides a portal to new physics. Further, knowledge of 
the neutron distribution helps improve the isovector sector of nuclear energy density functionals. In turn, refinements 
to the isovector sector translate into powerful constraints on the density dependence of the symmetry energy, and 
ultimately on the equation of state of neutron-rich matter. Finally, constraining the equation of state of neutron-rich 
matter is essential to fully capitalize on new discoveries in this new era of gravitational wave astronomy.

\begin{acknowledgments}
 JP thanks the Institute for Nuclear Theory at the University of Washington for its kind hospitality and 
 stimulating research environment. This material is based upon work supported by the U.S. Department of Energy 
 Office of Science, Office of Nuclear Physics under Award Number DE-FG02-92ER40750. 
\end{acknowledgments}
\vfill\eject

\bibliography{EWProbes.bbl}

\vfill\eject
\end{document}